\documentclass[english,twocolumn,notitlepage,color,superscriptaddress,psfig,showpacs,amsmath,amssymb,aps,prb,floatfix,reprint,epsf,nobibnotes,longbibliography]{revtex4-2}

\usepackage[utf8]{inputenc}
\usepackage{marvosym}
\usepackage[colorlinks=true,citecolor=blue]{hyperref}
\usepackage{graphicx,subfigure}     
\usepackage{dcolumn}        
\usepackage{bm}             
\usepackage{bbm}
\usepackage{verbatim}
\usepackage{epsf}
\usepackage{amssymb,stmaryrd}   
\usepackage{color}
\usepackage{float}
\usepackage{epstopdf}
\usepackage{natbib}
\usepackage{amsmath}
\usepackage{mathrsfs}
\usepackage{xspace}
\usepackage{placeins}
\usepackage{babel}
\usepackage{color,graphicx}
\usepackage{caption}
\graphicspath{ {images/} }  
\usepackage{wrapfig}
\usepackage{mathtools}
\usepackage{tikz}  
\usetikzlibrary{arrows,shapes,trees,positioning}  
\usepackage{comment}
\usepackage{siunitx}

\usepackage{appendix}
\usepackage{textcomp}
\usepackage[mathlines]{lineno}    
\usepackage{soul}
\setstcolor{red}
\usepackage[ignoreunlbld,norefs,nocites]{refcheck}
\usepackage{ulem}
\usepackage{multirow}
\hypersetup{
	colorlinks=true,
	linkcolor=blue,
	citecolor=blue,
	urlcolor=blue,
}

\begin{document}


\title{Adaptive Synaptogenesis  Implemented on a Nanomagnetic Platform}

\author{Faiyaz Elahi Mullick}
\email{fm4fv@virginia.edu}
\affiliation{Department of Electrical and Computer Engineering, University of Virginia}

\author{Supriyo Bandyopadhyay}%
\email{sbandy@vcu.edu}
\affiliation{Department of Electrical and Computer Engineering, Virginia Commonwealth University}

\author{Rob Baxter}%
\email{rbaxter37@gmail.com}
\affiliation{Department of Neurosurgery, University of Virginia}

\author{Tony J. Ragucci}%
\email{tony.ragucci@drs.com}
\affiliation{Leonardo DRS}

\author{Avik W. Ghosh}%
\email{ag7rq@virginia.edu}
\affiliation{Department of Electrical and Computer Engineering, University of Virginia}
\affiliation{Department of  Physics, University of Virginia}

\begin{abstract}
The human brain functions very differently from artificial neural networks (ANN) and possesses unique features that are absent in ANN. An important one among them is ``adaptive synaptogenesis'' that modifies synaptic weights when needed to avoid {\it catastrophic forgetting} and promote lifelong learning. The key aspect of this algorithm is supervised Hebbian learning, where weight modifications in the neocortex driven by temporal coincidence are further accepted or vetoed by an added control mechanism from the hippocampus during the training cycle, to make distant synaptic connections highly sparse and strategic. In this work, we discuss various algorithmic aspects of adaptive synaptogenesis tailored to edge computing, demonstrate its function using simulations, and design nanomagnetic hardware accelerators for specific functions of synaptogenesis.
\end{abstract}

\maketitle

\section{Introduction}

The need for brain-inspired computing far beyond conventional artificial neural networks (ANN) is a pressing one. Conventional mahine learning (ML) based on Deep Neural Nets (DNN) is ill-suited for the rapid growth in edge intelligence owing to the considerable cost of wiring, exorbitant energy demands,  and a limit on memory resources. In ANN/DNN, new sensory data are learned independent of past  history, requiring added hardware and costly synaptic interconnects, with an exploding area footprint and energy budget. In edge environments requiring on-chip computing, such as autonomous robots exploring mines, battlefield vehicles navigating enemy terrain, underwater drones or Mars rovers, robots must analyze rapidly varying situations with very limited memory resources and energy budgets, without relying on a cloud that is either unavailable or unreliable in the face of security breaches. In 3D Autonomous Simultaneous Localization and Mapping (ASLAM) for mobile robotics, for instance, robot actions need to be calculated on the go offline, but this is traditionally done by deleting out-of-date data using a delayed nearest neighbor data association strategy \cite{Hans,Mei,Lluvia}. Going beyond such a local adaptive mapping and navigation, one step at a time, becomes completely prohibitive over an expanded time horizon. 

In a distributed neural net, the challenge with
learning sequential data is called the {\it{stability-plasticity dilemma}}, the choice between integrating new knowledge vs remembering previously acquired knowledge. The conventional way to deal with this is to reduce overlap among the stored internal representations, for example by using sparse or interleaved learning. Connection networks must then absorb new inputs and adjust synaptic weights by {\it small increments}, thereby preventing sequential learning. 
Attempts to address this challenge have mostly been made in software, for instance, a Fahlman offset \cite{Mermillod} in the derivative of the sigmoid function in backpropagation to avoid entrenchment in its flat parts.  Intel's recent AI robot \cite{hajizada} relied on a learning phase where prototype data are moved around but not erased in feature space, punishing the wrong category or rewarding the right category, and only allocating new resources if the error persists and an unknown category is thereby identified. Extending learning to a practical hardware environment, with acceptable Size Weight and Power (SWaP) is still an unrealized target.
    \begin{figure*}[!hbt]
\includegraphics[width=\textwidth]{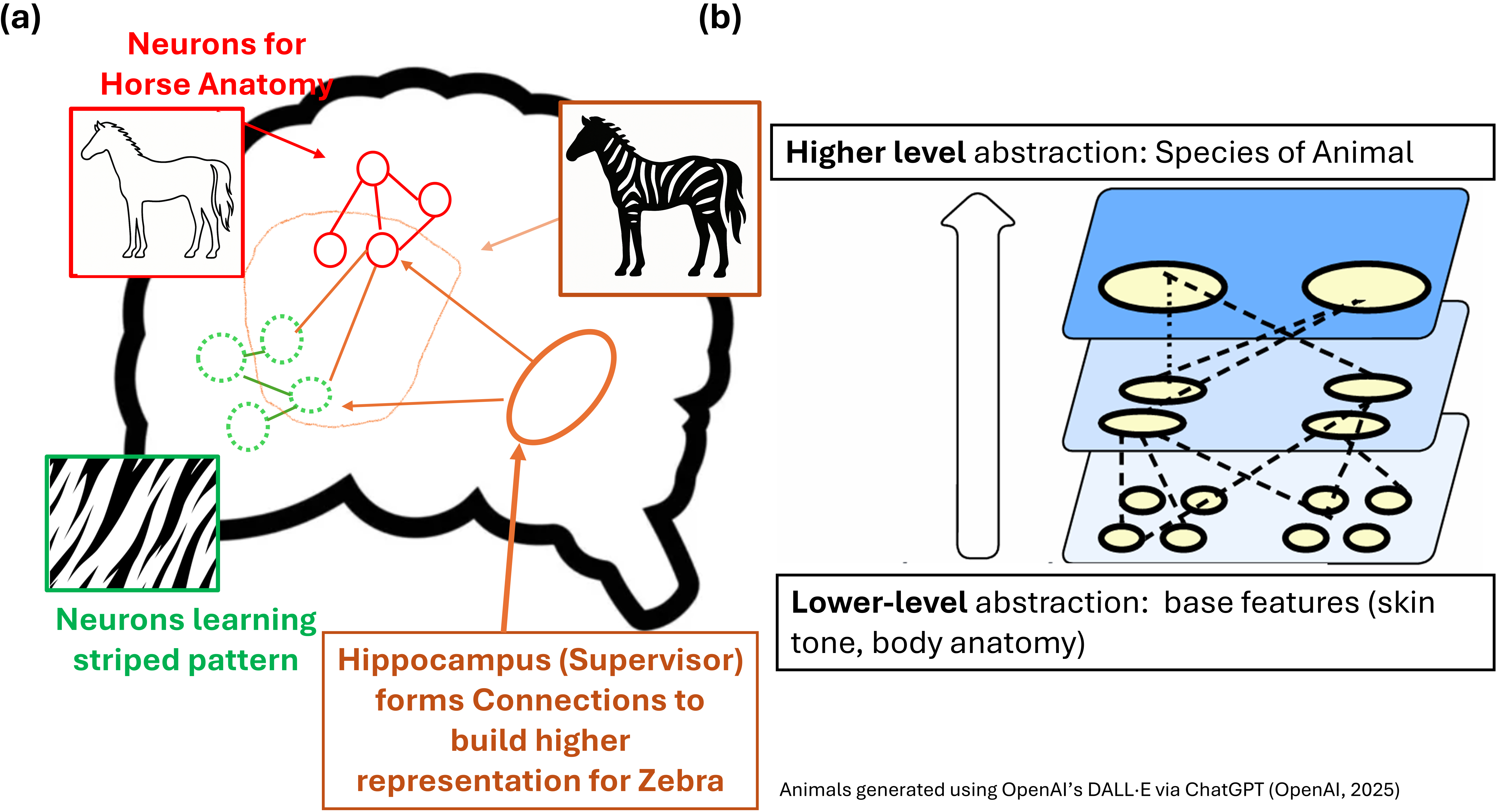}
\captionsetup{width=\columnwidth}
    \caption{(a) lower level representations of stripes and animal anatomy learnt by groups of neurons that will get wired together to (b)  represent a higher level abstraction such as an animal species of zebra}
\label{fig:horse}
\end{figure*}
In contrast, human cognitive abilities provide a remarkable example of how the brain encodes sequential learning very efficiently with a fixed memory bank, avoiding {\it{\bf catastrophic forgetfulness}} (sequential data over-write) with {{incremental}} and contextual {{learning}}. The evolved trick is an over-riding hippocampus  (HC) that acts as a memory support structure for a severely  resource-constrained neocortex (NC), with dense local but very sparse, one in a million, non-local connectivity  that mediates perceptions and decisions. In contrast to brain development which proceeds from sensory, up through the cortical hierarchy (i.e., bottom-up), later learning proceeds in a top-down manner directed by the hippocampal system, a dense recurrently connected generative network with the ability to process and correlate long term multi-dimensional sequences \cite{Levy1996,August1999} to build a sophisticated “World Model” for the organism. 
    In addition to the hardware support structure of a dual memory, the brain also uses algorithmic techniques to encode episodic memory by hierarchically implementing depths of representation. Big ideas are built out of association between contexts, event histories and multi-sensory microscopic details; for instance reframing a `zebra' as a `striped horse'.(Fig.~\ref{fig:horse}).
\\
\indent {\it{The process of having a separate dense hippocampal network directing a sparser neocortical network allows us to avoid catastrophic forgetfulness and compose coherent long-term semantic memory {\bf (lifelong learning)}} from short, fragmented, conditionally independent, episodic events and their intermediate representations.} 

\section{Brain inspired learning models}
The goal of this section is to develop a mathematical model for the hippocampal-neocortical (HC-NC) interaction that allows a compute engine to maintain sparsity while encoding episodic associative memory - i.e., memory encoded as a sequence of time-correlated (but not necessarily causal) events. The key part of this algorithm is the employment of a digital synaptogenesis step - namely, the creation or removal (one or zero) of primarily distal synaptic connections, on top of an analog weight modification step, where the synaptic weights in the NC are increased when they co-fire in a Hebbian format (`neurons that fire together wire together'). In contrast to conventional Hebbian networks which are unsupervised, in this model the Hebbian coefficient is a probabilistic random variable, and on top of that the HC exercises an added supervisory control through digital synaptogenesis. In other words, this process is {\it{partially supervised}}.

We will introduce the mathematical algorithms in this section. In section III, we will discuss the implications of these algorithms with a model problem, while in section IV we will explore a potential hardware framework to realize this architecture on-chip using compact voltage tunable magnetic neurons. 

\subsection{Hebbian Learning}
\begin{figure}[!hbt]
    \includegraphics[width=0.5\textwidth]{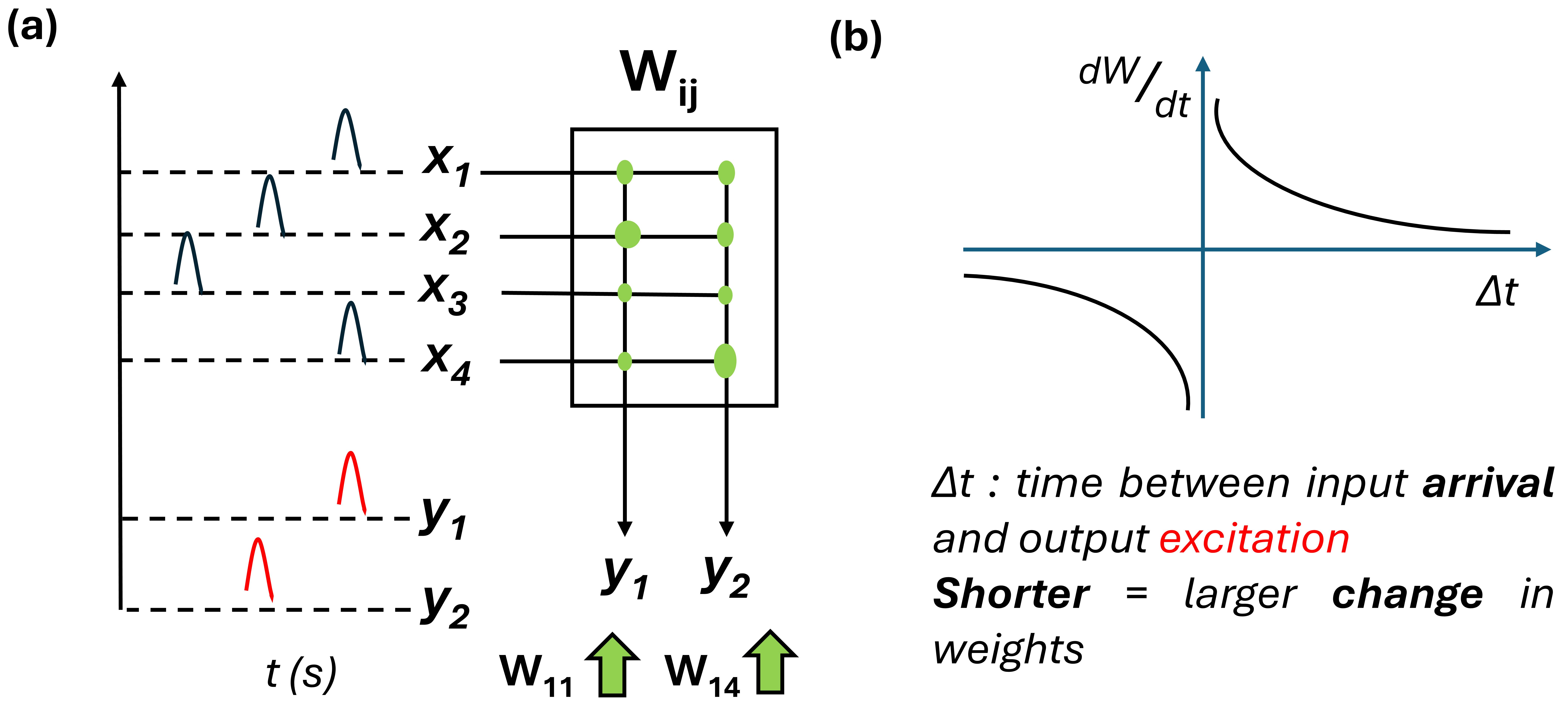}
    \captionsetup{width=0.8\linewidth}
    \caption{(a) Hebbian learning principle, $y_{1}$ firing just as inputs arrive at $x_1$ and $x_4$ implies weights $W_{11}$ and $W_{14}$ must be updated since Hebbian employs (b) spike timing dependent plasticity where a smaller $\Delta t$ causes larger change $\Delta W$.}
    \label{fig:hebbian_learning}
\end{figure}

The most widespread synaptic modification rule for biological models of the brain employ Hebbian learning, where weights are updated based on the time proximity of neuronal firing (Fig.~ \ref{fig:hebbian_learning}). Let us start with simple Hebbian neural network $H$, and define the main variables. Let the input vector  be $X=\{x_{1},x_{2},x_{3}, ...,x_{n}  \}$ where $x_{i}$ can be any analog value between 0 and 1. We do this because conventional datasets are continuous such as temperature readings from a factory sensor, pixel intensity in a 2D image, current values from a photo-detector etc. and are generally normalized between 0 and 1. Alongside this, we also have a set of dimensionless weights $W=\{ w_{11},w_{12},...w_{mn}\}$ where $w_{ij}$ is once again a continuous value between 0 and 1. For each input vector \textbf{X}, an output vector $Y=\{y_{1},y_{2},y_{3}, ...,y_{m}  \}$ exists in the training dataset where the definition of \textbf{Y} depends on the nature of the task:

\begin{itemize}
    \item Regression : Since the objective is to predict continuous values, each $y_{j}$ represents a real-valued  prediction corresponding to a continuous variable and $m$ represents how many outputs the model will learn to predict. As such, the outputs are typically normalized to fall within 0 and 1. Formally, for each input \textbf{X}, the output is:
    \begin{equation}
        Y = H(X; W)
    \end{equation}
    Where, the Hebbian model \textbf{H} will predict \textbf{Y} using \textbf{W} and \textbf{X}.
    
    \item Classification : Here the output represents a probability distribution over $m$ discrete classes and each $y_{j}$ indicates the predicted probability that the input \textbf{X} belongs to class $j$ and the vector satisfies the constraint:
    \begin{equation}
        \sum_{j=1}^{m} y_j = 1
        \label{eqn:classification_label}
    \end{equation}
    This ensures that the output forms a valid probability distribution. The predicted class is typically chosen as the index with the highest probability (magnitude):
    \begin{equation}
        j_{max} = \arg\max_j y_j
    \end{equation}
    The ground truth label (original labels from the dataset) are represented using one-hot encoding such that the predicted label \textbf{Y*} will have elements:
    \begin{equation}
        y^*_j = 
\begin{cases}
1, & \text{if } j = j_{max}\\
0, & \text{otherwise}.
\end{cases}
    \end{equation}
    Let us provide a concrete example. Let us say you have a 2D image dataset of cats, dogs and deer. Your dataset input vector size will be determined by the number of pixels each image has, but the size of your output vector will be 3 since there are three labels. The one hot encoded label for each would be 100, 010 and 001 respectively. This means $Y=\{y_{1},y_{2},y_{3}\}$ where $y_{j}$ follows (\ref{eqn:classification_label}).
    In actual practice, the predicted output $y_{j}^*$'s will not exactly match the one hot encoded labels. Instead, it will be a distribution. For example, if the input image was that of a cat whose label is 100, the output \textbf{Y} might look like $y_{j} =\{0.8,0.15,0.05\}$ where the maximum of 0.8 indicates the model predicted the image is a cat with 80 \% confidence.
\end{itemize}

 Now, a Hebbian model does not use a loss function $L$ like a traditional network and instead the extent to which we will change any weight $w_{ij}$ depends on the time difference $\Delta t$ between an input $x_{i}$ arriving and the excitation of any output $y_{j}$(Fig.~\ref{fig:hebbian_learning}(b)). This makes our Hebbian weight update rule at any time $(t)$ as:
\begin{equation}
\Delta w_{ij}(t) = 
\begin{cases}
x_{i}(t). y_{j}(t) . A_+ .e^{-\Delta t / \tau_+}, & \text{if } \Delta t > 0 \\
x_{i}(t). y_{j}(t) . A_- .e^{\Delta t / \tau_-}, & \text{if } \Delta t < 0
\end{cases}
    \label{hebbian}
\end{equation}
where $A_+ \text{ and } A_-$ are scaling factors (learning rate), $\Delta t$ is the time between an input $x_{i}$ arriving and an output $y_{j}$ firing, and $\tau_+,\tau_-$ are time constants for potentiation and depression, respectively. This continuous-time model captures the idea that causally aligned spikes (input before output) lead to synaptic strengthening, while anti-causal spikes (input after output) lead to weakening. 

For our computation, we discretize time into $n$ discrete timesteps $t_{n}$ such that inputs and outputs are sampled in intervals of $\delta t$ obeying $t_{n} = n \cdot \delta t$ where $n \in \mathbb{N}$. We also simplify (\ref{hebbian}) by assuming both $x_{i}$ and $y_{j}$ must be non-zero at the same $t_n$ (i.e., replace the exponents with a `box' function, and focus here on potentiation alone). This means we assume $\Delta t \rightarrow 0$ i.e. weight update will only happen at $t_n$:
\begin{equation}
\Delta w_{ij}(t_n) =
\begin{cases}
\epsilon \cdot x_i(t_n) \cdot y_j(t_n) , & \text{if } x_i(t_n), y_j(t_n) > 0 \\
0, & \text{otherwise}
\end{cases}
\label{eq:discrete_hebb}
\end{equation}
where $\epsilon$ combines both $A_+, A_-$ into a single learning rate, and $\Delta W_{ij}(t_n)$ only updates at timestep $t_n$ when both input and output are non-zero at $t_n$.
We further modify (\ref{eq:discrete_hebb}) by adding an upper bound
\begin{equation}
\Delta w_{ij}(t_n)= \epsilon \Biggl(x_{i}(t_n)-w_{ij} (t_n)-E(x)\Biggr)y_{j}(t_n),
\label{weight_modification}
\end{equation}
which ensures that the weights decay to zero when the feature is learned completely and fluctuations of the input $x_i$ around the expectation $E(x)$ minimize. Also, the weights will decrease in magnitude anytime an output fires in the absence of an input i.e. a `false' positive. We can then find the excitation as :
\begin{equation}
    y_{j} = f\left( \frac{\textbf{W}^{T}_{j} \textbf{X}}{\sum_{i}^{n} w_{ij}+A}\right),
    \label{activation}
\end{equation}
where we divide by the sum of lateral weights plus a constant $A$ between 0 and 1 to constrain the excitation. We use the factor $A$ for further control on this inhibition since initially the lateral sum $\sum^n_i w_{ij}$ might not be large enough to control the output.

We then pass the output $y_{j}$ through an activation function $f$ where $f$ can be \cite{hendrycks2016gelu} a perceptron, sigmoid, tanh, ReLU, GeLU etc. and the nature of $y_{j}$ will depend on the function being used. For example, with tanh, $y_{j}$ is in the range of $-1$ and $+1$ while with ReLU $y_{j} =  \frac{\textbf{W}^{T}_{j} \textbf{X}}{\sum_{i}^{n} w_{ij}+A}$ if $y_{j}> \theta_{y}$  (a given threshold on the output), otherwise 0.
\subsection{Synaptogenesis and Shedding}
 Whenever new information is presented to biological brains, they either modify existing neural connections or form new ones. While classical artificial neural networks also modify connections, where they depart from biology is the formation of new connections. In ANN, all neurons are pre-connected and their weights are imposed during a training step, while in biological networks, many (most) synaptic connections are not even eligible. This biological phenomenon is called {\it synaptogenesis}. By using a combination of Hebbian weight modification and synaptogenesis, biological brains learn and remember new information.
 
\begin{figure}[!hbt]
    \centering
    \includegraphics[width=0.5\textwidth]{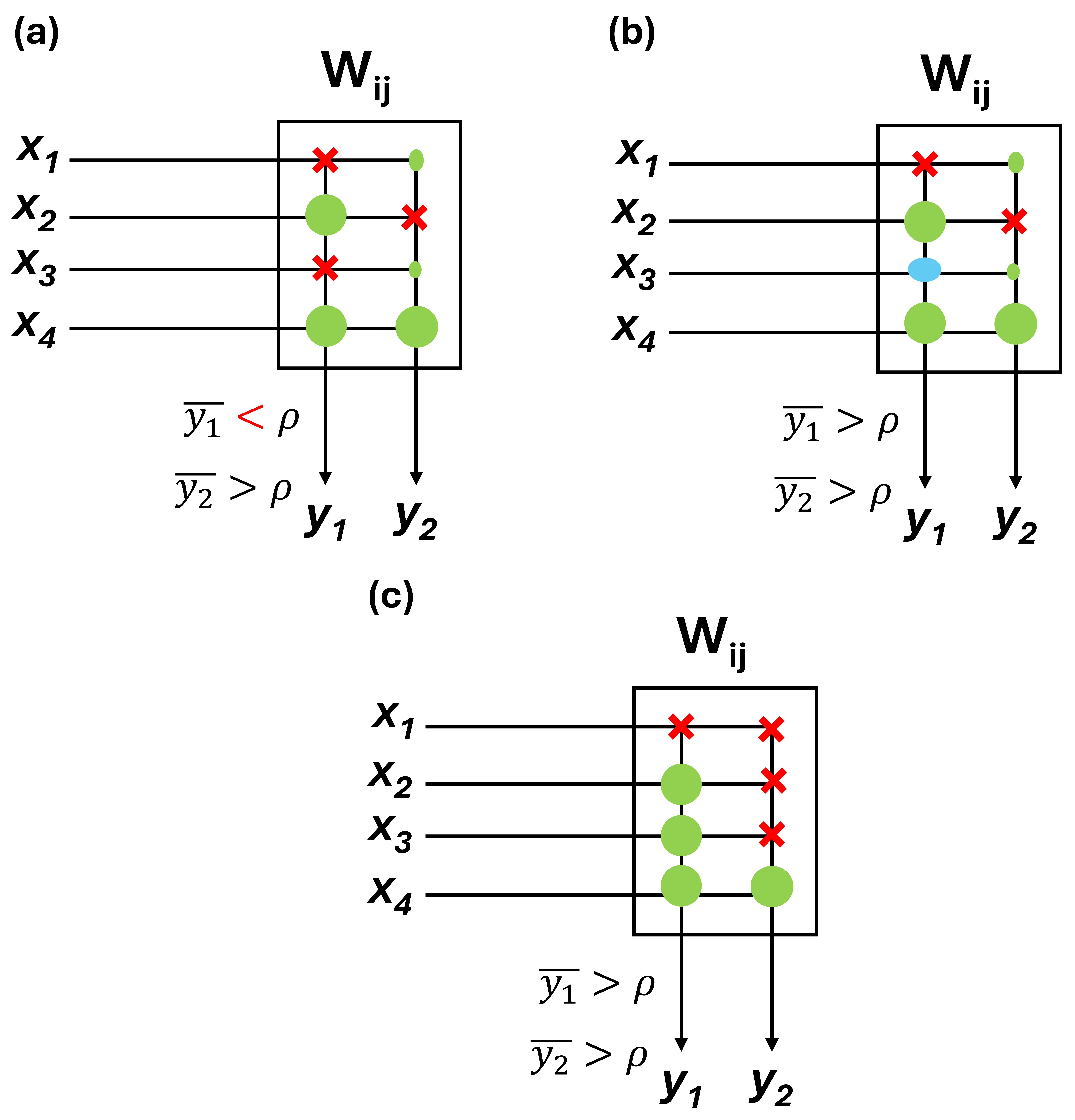}
    \captionsetup{justification=justified, width=0.8\linewidth}
    \caption{(a) Initial state where inputs are not stimulating $y_{1}$ enough, so its firing rate is low. 
    (b) Synaptogenesis forms a new connection, increasing the firing rate above the threshold. Also, notice the smaller weights connected to $y_{2}$. 
    (c) Low-value weights are 'shed,' and the firing rate remains unaffected, indicating they were not contributing to the activation of $y_{2}$.}
    \label{fig:synapto_illustration}
\end{figure}

 One major modeling question for synaptogenesis is what guides the formation of new connections? While many complex methods exist, in our approach \cite{Baxter2020ConstructingMechanisms} we chose to use the average firing rate of a neuron to guide synapse formation. We define the calculation of the average firing rate using the following equation:
     \begin{equation}
        \bar{y}_{j}(t_n) = (1-\alpha) \bar{y}_{j}(t_n-1) + \alpha y_{j}(t_n-1)
        \label{avg_rate}
    \end{equation}
where the current average firing rate $\bar{y}_{j}(t)$ is a running average calculated using the previous average $\bar{y}_{j}(t_n-1)$ and a `portion' of the current instantaneous (no bar) output $y_{j}(t_n-1)$ at each timestep ($t_n$). The variable $\alpha$ controls how much of the current output is used to update this running average.
 The key idea is that, for any output neuron $y_{j} = \sum_{i=1}^{n} w_{ij} x_i$ connected to \textit{n} input neurons $x_{i}$ via synapses $w_{ij}$ (Fig. \ref{fig:synapto_illustration}a), every time inputs excite $x_{i}$, they should end up firing $y_{j}$. We expect  $y_{j}$ to fire consistently with the frequency of inputs presented to $x_{i}$. However, sometimes the inputs might not be able to generate a strong enough excitation on the output. This is when synaptogenesis occurs as shown in Fig. \ref{fig:synapto_illustration}(a) where new synapses are added if the average firing rate falls below some threshold $\rho$. Synapses are formed followed by modification using (\ref{weight_modification}). 
 
 Once synapses are added and modified, a final scan is done to check for any weights that are contributing too little to the excitation to incorporate a synaptic shedding step to reinforce sparsity. This is as simple as having a shedding threshold, so that any weights falling below  are removed from $W$.
 
 Biological brains do not simply form connections immediately after a neuron fails to excite properly, they also consider the cost of such an action. They maintain sparsity and efficiency by only forming connections when a neuron is constantly excited and fails to activate. This means that there is a probabilistic aspect to synaptogenesis -- the probability of the brain forming new connection(s) is less (often much less) than unity. To incorporate this into Hebbian learning, we first introduce an additional matrix of 0-s and 1-s called the {\it connection matrix} $C_{ij}$. This matrix tracks which inputs $x_{i}$ are connected to each output $y_{j}$. This converts equations (\ref{weight_modification}), (\ref{activation}) into:
 \begin{itemize}
    \item The modified weight update rule:
    \begin{equation}
        \Delta w_{ij} = \epsilon  C_{ij}  \Biggl(x_{i} - w_{ij} - E(x)\Biggr)  y_{j}
    \end{equation}
    
    \item The new output excitation equation:
    \begin{equation}
        y_{j} = f\left(\frac{\textbf{C}^{T}_{j} \textbf{W}^{T}_{j} \textbf{X}}{\sum_{i}^{n} w_{ij} + A}\right)
    \end{equation}
\end{itemize}
 The connection matrix $C_{ij}$ responsible for tracking synaptogenesis ensures that only connected synapses update their weights.
\begin{figure*}[!hbt]
    \centering
    \includegraphics[width=\textwidth]{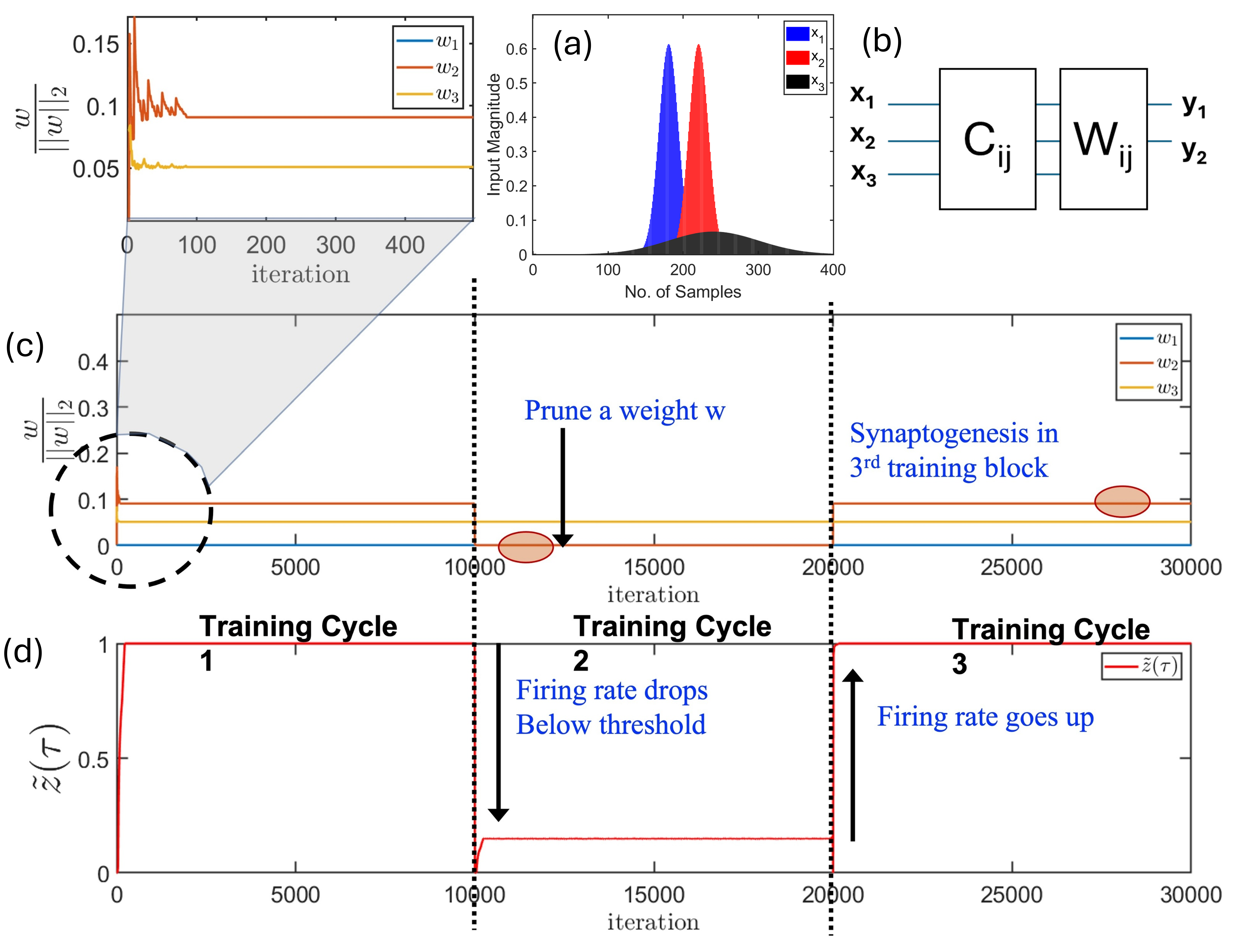}
    \captionsetup{justification=justified, width=0.9\linewidth}
    \caption{(a) Distribution of input dataset (500 points) run on (b) model with two outputs. (c) We monitor a few randomly selected weights over iterations (each iteration sends one datapoint across 60 epochs. (d) Pruned weight reduces firing rate which causes model to add weights until firing rate meets threshold}
    \label{fig:synapto_demo}
\end{figure*}

 We now codify probabilistic synaptogenesis in a set of equations that control the elements of $C_{ij}$ based on the average firing rate of each output neuron.
\begin{itemize}
    \item For any input $x_{i}$ exciting outputs $y_{j}$, we find the running average firing rates of $y_{j}$ using (\ref{avg_rate}).
    
    \item For each connection between $x_{i}$'s and $y_{j}$'s, we find the associated probabilities $p_{ij}$ as
    \begin{equation}
        p_{ij} = \gamma (1-C_{ij}) a_i 
    \end{equation}
    where $\gamma$ is the synapse formation rate and lies between 0 and 1. A value of 1 would imply fully deterministic  synapse formation and a value of 0.02 would imply roughly 2\% connectivity. We use a variable $a_{i}$ to check for the presence of input $x_{i}$,    i.e., $a_{i} = 1$ for $x_{i} >0$ amd 0 otherwise. This allows the probability to only be computed when an input is present i.e. non-zero.
    \item We update the connection matrix according to the following rule.  For each $p_{ij}$ sampled from a uniform distribution $B \sim U(0,1)$ we change $C_{ij}$ thus:
    \begin{itemize}
        \item if $p_{ij} > B_{ij}$, $C_{ij} = 1$ and $w_{ij} = W_{o}$, where $W_{o}$ is the initial weight value for any new connection (this is chosen based on dataset distribution)
        \item if $p_{ij} < B_{ij}$, then $C_{ij}$ remains unchanged.
    \end{itemize}
    The random variable \textbf{B} allows synaptogenesis to be probabilisic and $\gamma$ controls how frequently synaptogenesis can occur. 
    
    Finally shedding occurs near the end:
    \item if any $W_{ij}<W_{shed}$, $C_{ij} = 0$, where $W_{shed}$ is the threshold of cutting off redundant weights.
\end{itemize}
To recap, $W$ is an analog weight that attempts to update using coincidence, while $C$ is an overriding digital enabling circuit that enables or vetoes these attempted modifications. In biological systems, these operate on different time-scales or epochs. In our model, we assume they are evaluated concurrently at each step. 

The combined set of rules comprehensively describes how {\it synaptogenesis} occurs. 

\section{Model Simulations}

\subsection{Demonstration of Synaptogenesis}
To demonstrate how the model creates new connections, we set up a simple experiment. Let us use a classification example. At first, (Fig.~ \ref{fig:synapto_demo}(a)) we generate a synthetic distribution with three inputs $x_{1}$, $x_{2}$ and $x_{3}$  (500 data points)  for a model with two outputs $y_{1}$, $y_{2}$ which can be either 0 or 1 i.e. a classification scenario. 
We are simply trying to define a dataset that has only two possible labels for three different inputs. For our activation function we will use the perceptron which  makes $y_j = 1$ if $y_{j}> \theta_{y}$. We initialize the run with $\gamma = 1$, $\epsilon = 0.05$, $\alpha =0.01$ and activation threshold $\theta_{y} = 0.02$ on $y_j$ and train the model for a total of 60 epochs, where each epoch is defined as the time required to train the model over the entire dataset once. Whenever a new epoch is started, the input dataset is randomized and the model is trained again. For 500 data points we would be training the model for a total of 30,000 time steps, while monitoring three randomly chosen weight values. In the middle of the run, we purposefully prune one weight and notice the average firing rate of one of the outputs to fall (Fig. \ref{fig:synapto_demo}(d)) which causes the model to form a connection immediately to recover the firing rate.

\begin{figure}[!hbt]
    \centering
    \includegraphics[width=0.5\textwidth]{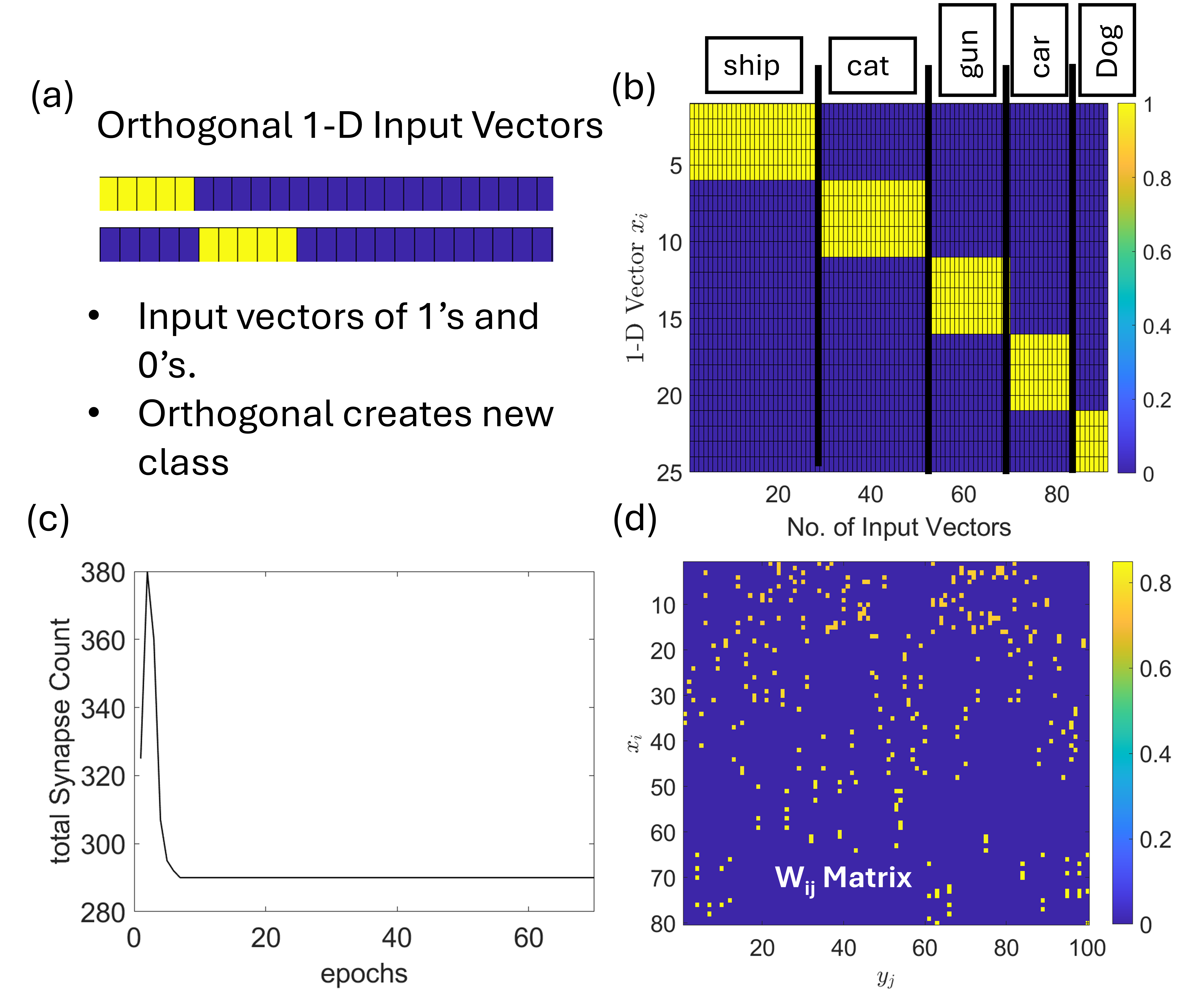}
    \captionsetup{justification=justified, width=0.8\linewidth}
    \caption{(a) 1-D generated binary vectors where orthogonality ensures distinct classes. (b) Dataset used with 5 classes, variation in no. of datavectors (c) Model optimizes number of synapses after a few epochs (d) Sparse $W_{ij}$ matrix with higher number of points near the top since the associated inputs were excited the most.}
    \label{fig:unsupervised_model}
\end{figure}

Now that we have explained synaptogenesis in a toy model, let us model a higher dimensional data set to see how we can generate sparse connections in the weight matrix. We generate a synthetic data set of 1-D orthogonal vectors (Fig. \ref{fig:unsupervised_model}(a)), where each unique orthogonal vector belongs to a distinct class (Fig. \ref{fig:unsupervised_model}(b)). We run the model according to the parameters in Table \ref{tab:unsupervised_parameters}. Here we see how initially the connection count climbs up until the model finally sheds to an optimal number of connections in Fig. \ref{fig:unsupervised_model}(c), resulting in a sparse final weight matrix (Fig. \ref{fig:unsupervised_model}(d)).
\begin{table}[h]
    \centering
    \begin{tabular}{|c|c|c|}
        \hline
        \textbf{Name}  & \textbf{Value}  \\ \hline
        synapse formation rate  $\gamma$ & 0.001    \\ \hline
        moving average window  $\alpha$ & 0.001    \\ \hline
        learning rate  $\epsilon$ & 0.05   \\ \hline
        initial weight  $W_{o}$ & 0.2   \\ \hline
        shed threshold  $W_{shed}$ & 0.01   \\ \hline
        epochs & 70    \\ \hline
        average firing rate $\rho$ & 0.1     \\ \hline
        inhibition factor A & 0.001   \\ \hline
        input neurons $x_{i}$ & 80    \\ \hline
        output neurons $y_{j}$ & 100    \\ \hline
    \end{tabular}
    \caption{Unsupervised Model Parameters}
    \label{tab:unsupervised_parameters}
\end{table}

\subsection{Supervised Synaptogenesis}
\begin{figure}[!hbt]
    \centering
    \includegraphics[width=0.5\textwidth]{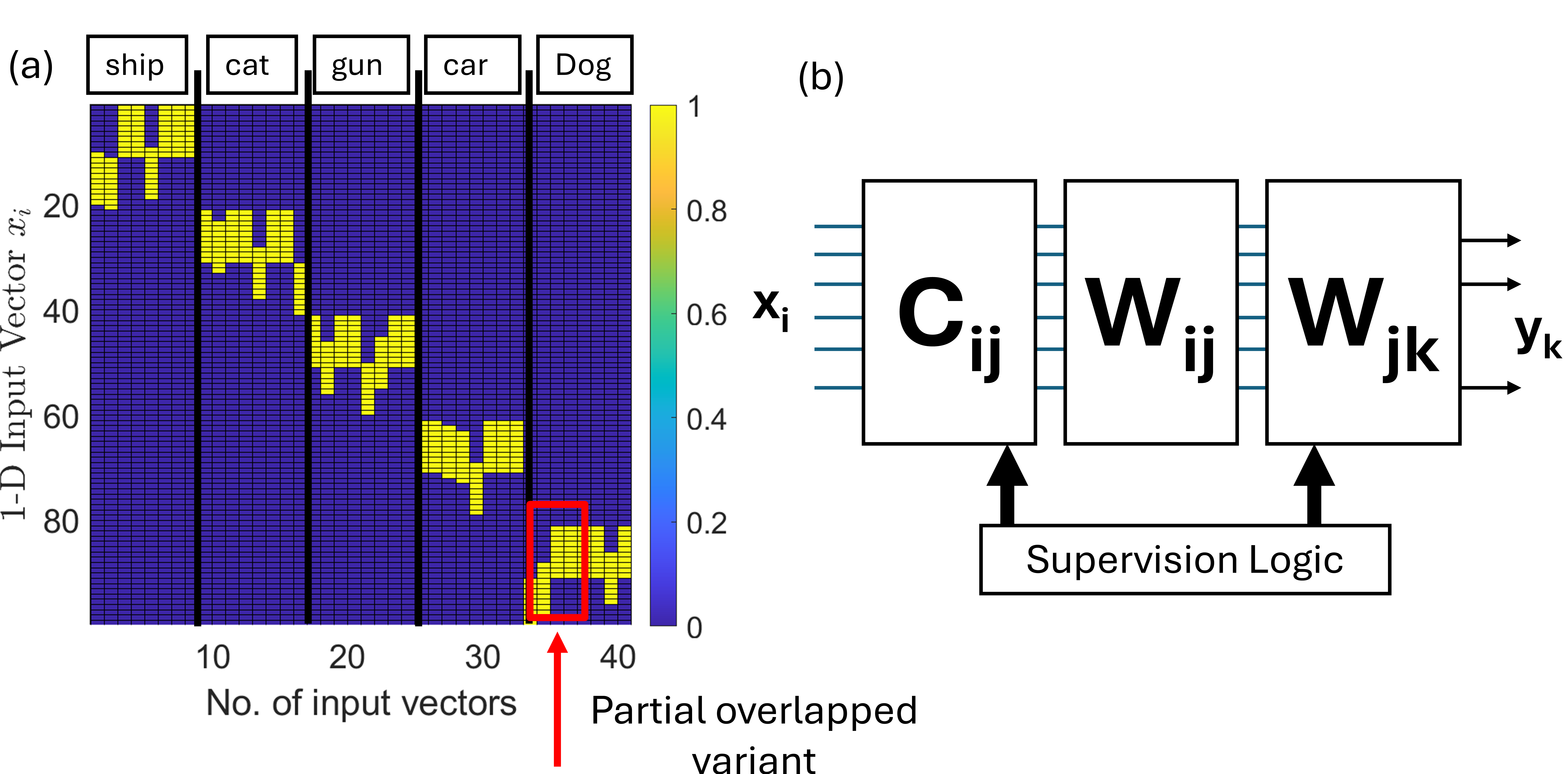}
    \captionsetup{justification=justified, width=0.8\linewidth}
    \caption{(a) Dataset with shared features (b) Supervised model}
    \label{fig:supervised_dataset}
\end{figure}

While an unsupervised approach allows us to see how the model is functioning on a fundamental level, it is difficult to implement this in most applications, specially when we have labeled data sets. This is because not all applications benefit from data driven analysis, some examples such as image classification for labeled data \textbf{require} some sort of supervision to reward/punish the model to learn properly. For this reason, we modify the synaptic rules so that a supervision signal \cite{Baxter2020ConstructingMechanisms} guides the formation of synapses to maximize training accuracy. This expands the model into two layers (Fig. \ref{fig:supervised_dataset}(b)).
We now run a simulation to show that in a supervised scene, we see neuron reuse. We set up a scenario where we start out with 5 categories like in Fig. \ref{fig:unsupervised_model}(a),(b) with 5 fully orthogonal unique vectors. We then generate variants with partial overlaps in each class shown in Fig. \ref{fig:supervised_dataset}(a). For our simulation, we:
\begin{itemize}
    \item Generate variants with partial overlaps in each class ensuring no interclass overlaps.
    \item Generate a total of 50 variants per class (overlap range 5\% to 40\%)
    \item Train a model with 0 variants, then 1 variant per class upto 50 variants per class. This means we have a total of 100 different models.
    \item Repeat training 50 times with different random seeds (for probabilistic synaptogenesis) for a total of 2500 models
    \item Analyze neuron count plots and test accuracy (80:20 split).
    \item compare accuracy to $k$-nearest neighbors for benchmarking
\end{itemize}
\begin{figure}[!hbt]
    \centering
    \includegraphics[width=0.5\textwidth]{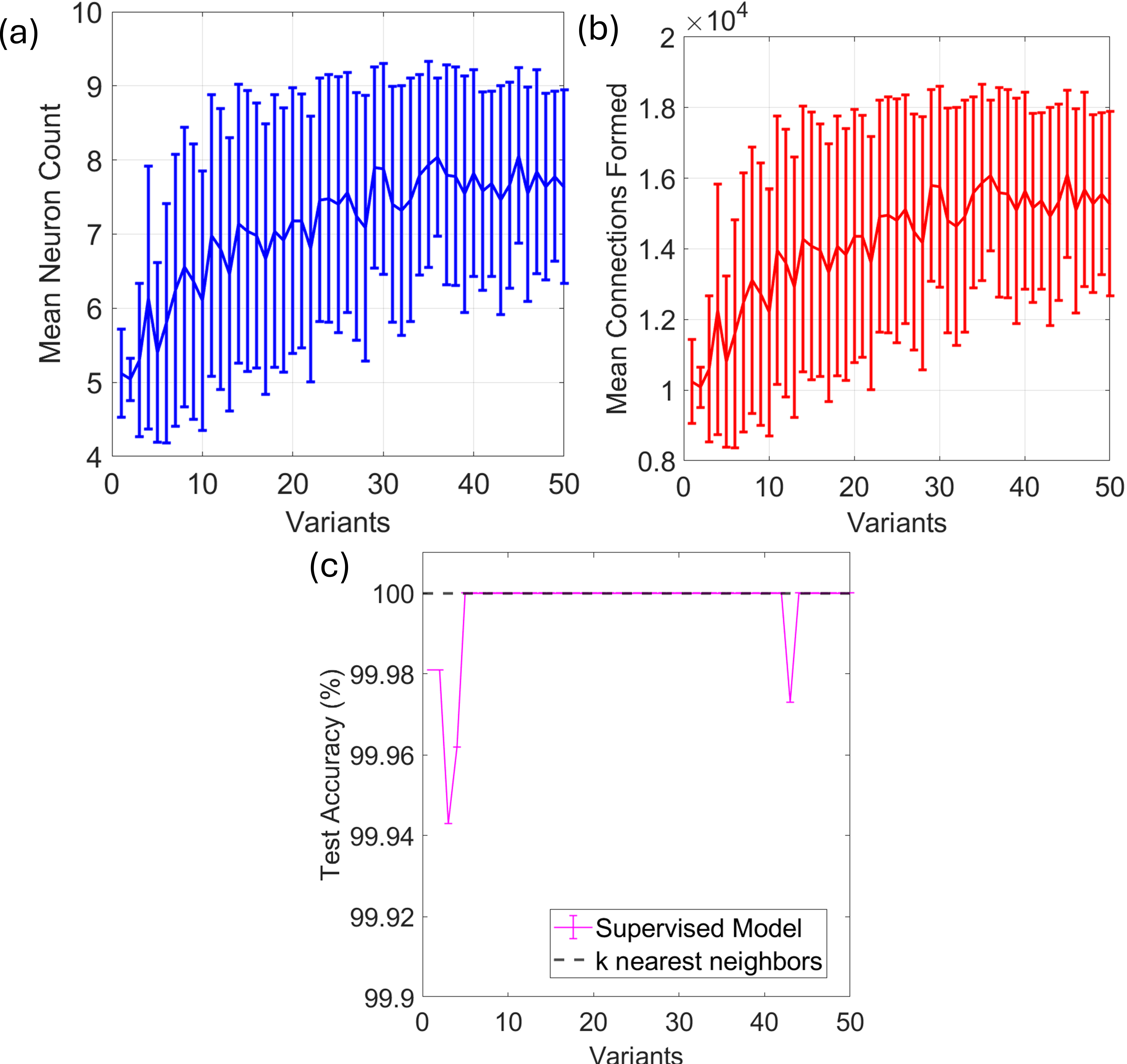}
    \captionsetup{justification=justified, width=0.8\linewidth}
    \caption{Average number of allocated(a) neurons (b) synapse connections employed during training of the supervised model using datasets with overlapping (shared features) and (c) their corresponding test set accuracy benchmarked against k nearest neighbors. We see neuron allocation 'plateau's as the shared features between inpput vectors allows re-use of neurons}
    \label{fig:supervised_results}
\end{figure}
Fig. \ref{fig:supervised_results} shows the results of our runs. We see an initial rise in neuron count with increase in variation. However this tapers off since the model no longer needs new neurons to reach maximum training accuracy. This shows that the dataset's shared features were learnt by the model and the model was able to re-use the same neurons. Accuracy is on par with $k$-nearest neighbors (kNN) for benchmarking reasons. We used sk-learn's packages for kNN. 
\begin{table}[h]
    \centering
    \begin{tabular}{|c|c|c|}
        \hline
        \textbf{Name}  & \textbf{Value}  \\ \hline
        synapse formation rate  $\gamma$ & 1    \\ \hline
        moving average window  $\alpha$ & 0.001    \\ \hline
        learning rate  $\epsilon$ & 0.05   \\ \hline
        initial weight  $W_{o}$ & 1   \\ \hline
        shed threshold  $W_{shed}$ & 0.01   \\ \hline
        epochs & 50    \\ \hline
        average firing rate $\rho$ & 0.1     \\ \hline
        inhibition factor A & 1   \\ \hline
        Number of trials  & 50   \\ \hline
        input neurons $x_{i}$ & 1   \\ \hline
        hidden neurons $y_{j}$ & 50    \\ \hline
        Output neurons $y_{k}$ & 5    \\ \hline
        Output activation threshold $\theta_{y}$ & 0.02    \\ \hline
    \end{tabular}
    \caption{Supervised Model Parameters}
\end{table}

\section{Hardware implementation of adaptive synaptogenesis}

We now discuss a set of hardware accelerators for adaptive synaptogenesis mainly comprising {\it nanomagnetic} devices for their superior energy-efficiency and non-volatility. 

The principal equation for weight modification in adaptive synaptogenesis is Equation (\ref{weight_modification}). It is clear from this equation that a hardware platform for adaptive synaptogenesis will require seven major components: (1) a device to measure the firing rate of a neuron, (2) a comparator for comparing the measured firing rate with a threshold, (3) an analog multiplier to multiply the analog (voltage or current) outputs of two neurons, (4) an analog subtractor to subtract one neuronal output from another, (5) analog (preferably non-volatile) weights, (6) a probability generator whose probability distribution can be tuned as desired, and (7) a neuron, such as a McCullough-Pitts type neuron or perceptron.

We will implement most of the hardware with magnetics, as opposed to electronics, primarily because magnetic devices are usually (not always) more {\it energy-efficient} than their electronic counterparts and they are {\it non-volatile}. Both attributes are very desirable for edge computing and/or computing in resource-constrained environments (deep space, underground or underwater, etc.) where energy sources are scarce and all data must be stored in-situ in non-volatile elements because the cloud is either unavailable (deep space) or unreliable (enemy territory). The best designed hardware will comprise mostly magnetic elements with a smattering of electronic elements for gain, interfacing, high-speed and error-resilience, if needed.

We describe below the design of the seven essential elements.

\subsection{The neuron}

We will implement a McCullough-Pitts neuron or perceptron with a straintronic spin neuron (SSN) \cite{ayan}. This model not only has its excellent energy efficiency and fast firing speed (sub-ns) but also keeps in line with our firing rate based synaptogenesis model. It relies on magnetization switching with electrically generated mechanical strain (hence the name ``straintronic''). 

\begin{figure}
\centering
\includegraphics[width=0.46\textwidth]{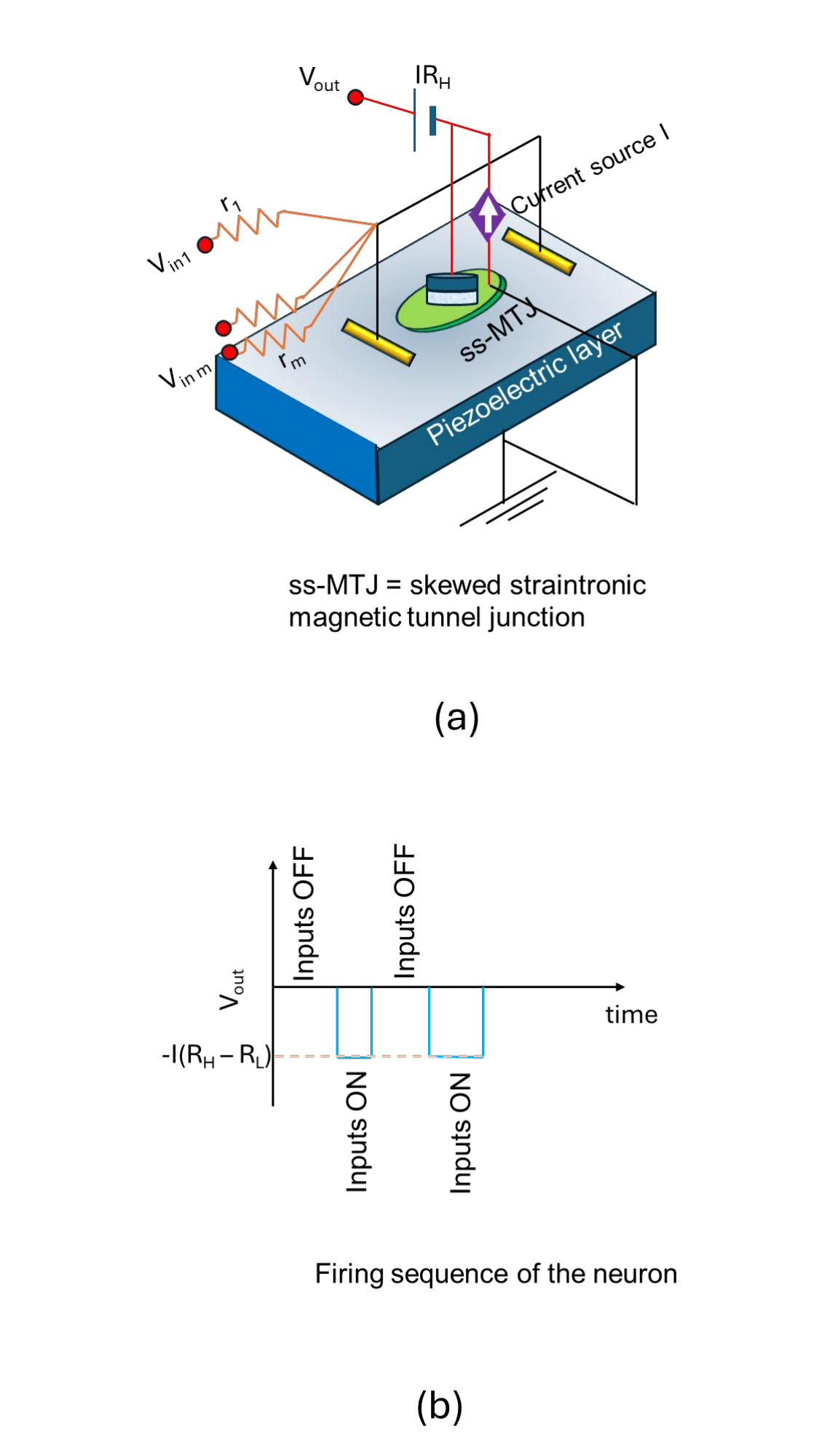}
\caption{\label{fig:SSN} (a) Schematic of a straintronic spin neuron. (b) The firing sequence of the neuron.}
\end{figure}

The device is shown in Fig. \ref{fig:SSN}(a) and consists of a `{\it skewed} straintronic magnetic tunnel junction' (ss-MTJ) which has elliptical magnetic hard and soft layers whose  major axes are {\it not} collinear (hence called `skewed'). The soft layer is in elastic contact with a piezoelectric thin film. The resistors $r_1 \cdot \cdot r_m$ encode  synaptic weights. Because of any residual dipole coupling between the hard and the soft layer, the ss-MTJ is in the high resistance state in the absence of inputs. When the weighted sum of the input voltages $V_{in1}, V_{in2} \cdot \cdot V_{in~m}$ exceeds a certain threshold, the strain generated in the piezoelectric (and transferred to the soft layer) flips its magnetization and abruptly switches the resistance of the ss-MTJ from the high value $R_H$ to the low value $R_L$, thereby switching the output voltage $V_{out}$  from 0 to $I \left ( R_L - R_H \right )$, where $I$ is a bias current driven through the ss-MTJ. The working of this neuron was explained in \cite{ayan} and hence omitted here. The output voltage $V_{out}$ is given by \cite{ayan}
\begin{equation}
    V_{out} = f \left ( \sum_i w_i V_{in~i} \right ),
\end{equation}
where $f$ is our perceptron function and 
\begin{equation}
    w_i = {{r_1 \parallel r_2 \parallel \cdot \cdot \cdot \parallel r_m}\over{r_1 \parallel r_2 \parallel \cdot \cdot \cdot \parallel r_m + r_i
    }}.
    \end{equation}

\begin{figure}
\centering
\includegraphics[width=0.46\textwidth]{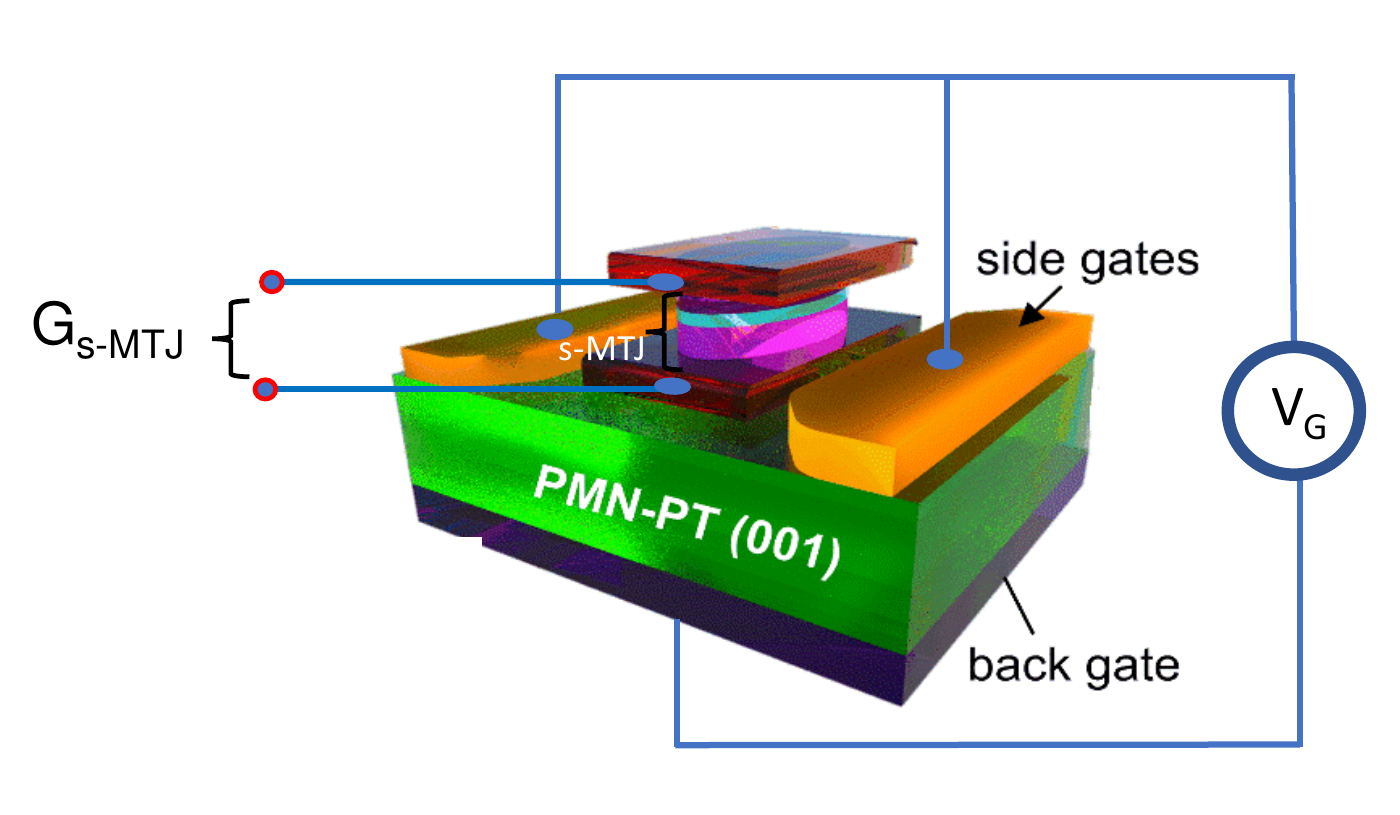}
\caption{\label{fig:s-MTJ} A straintronic magnetic tunnel junction (s-MTJ) reproduced with permission of the American Institute of Physics from \cite{zhao}. The gate voltage $V_G$ is applied across the piezoelectric PMN-PT between the (shorted) side gates and the back gate. This device exhibited a room-temperature tunneling magnetoresistance ratio (TMR) of slightly larger than 100\%.}
\end{figure}

\begin{figure}[!hbt]
 \vspace{0.1in}
\centering
\includegraphics[width=0.4\textwidth]{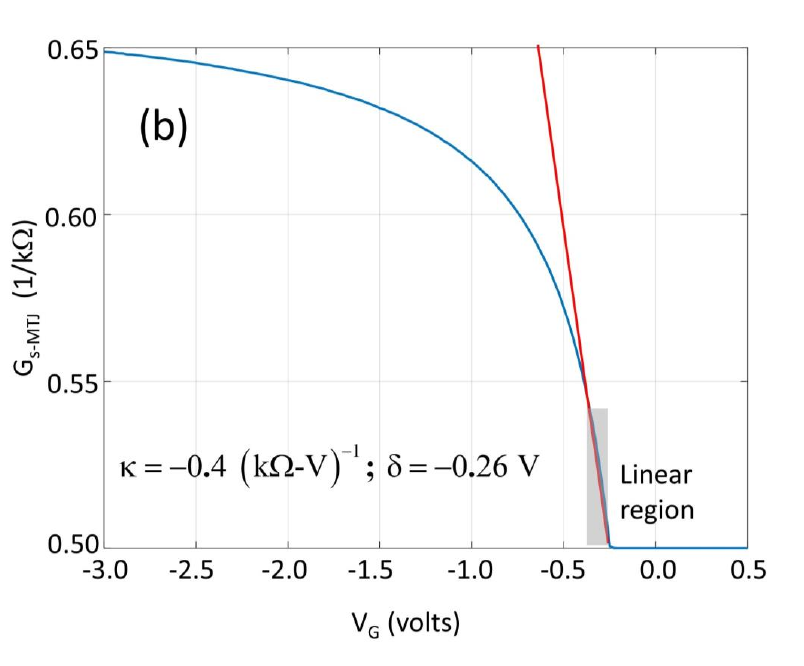}
\caption{\label{fig:transfer} The transfer characteristic of a straintronic magnetic tunnel junction computed with the use of stochastic Landau-Lifshitz-Gilbert simulation of the soft layer's magnetodynamics in the presence of 
    gate-voltage-generated strain at room temperature. This figure is reproduced from \cite{ieee} with the permission of the Institute of Electrical and Electronics Engineers. The parameters for the soft layer were major axis = 800 nm, minor axis = 700 nm, thickness = 2.2 nm, saturation magnetization $M_s$ = 8.5$\times$10$^5$ A/m, dipole coupling field $H_d$ = 1000 Oe, Gilbert damping constant = 0.1, saturation magnetostriction $\lambda_s$ = 600 ppm, Young's modulus $Y$ = 120 GPa, piezoelectric coefficient $d_{33}$ = 1.5$\times$10$^{-9}$ C/N and the piezoelectric layer thickness $t$ = 1 $\mu$m. The value of $R_{AP}$ was assumed to be 2 k$\Omega$ and the value of $R_P$ was 1 k$\Omega$.}
\end{figure}

The firing sequence of the neuron is shown schematically in Fig. \ref{fig:SSN}(b). 

\begin{figure*}[!hbt]
\vspace{-0.7in}
\centering
\includegraphics[width=0.8\textwidth,angle=270]{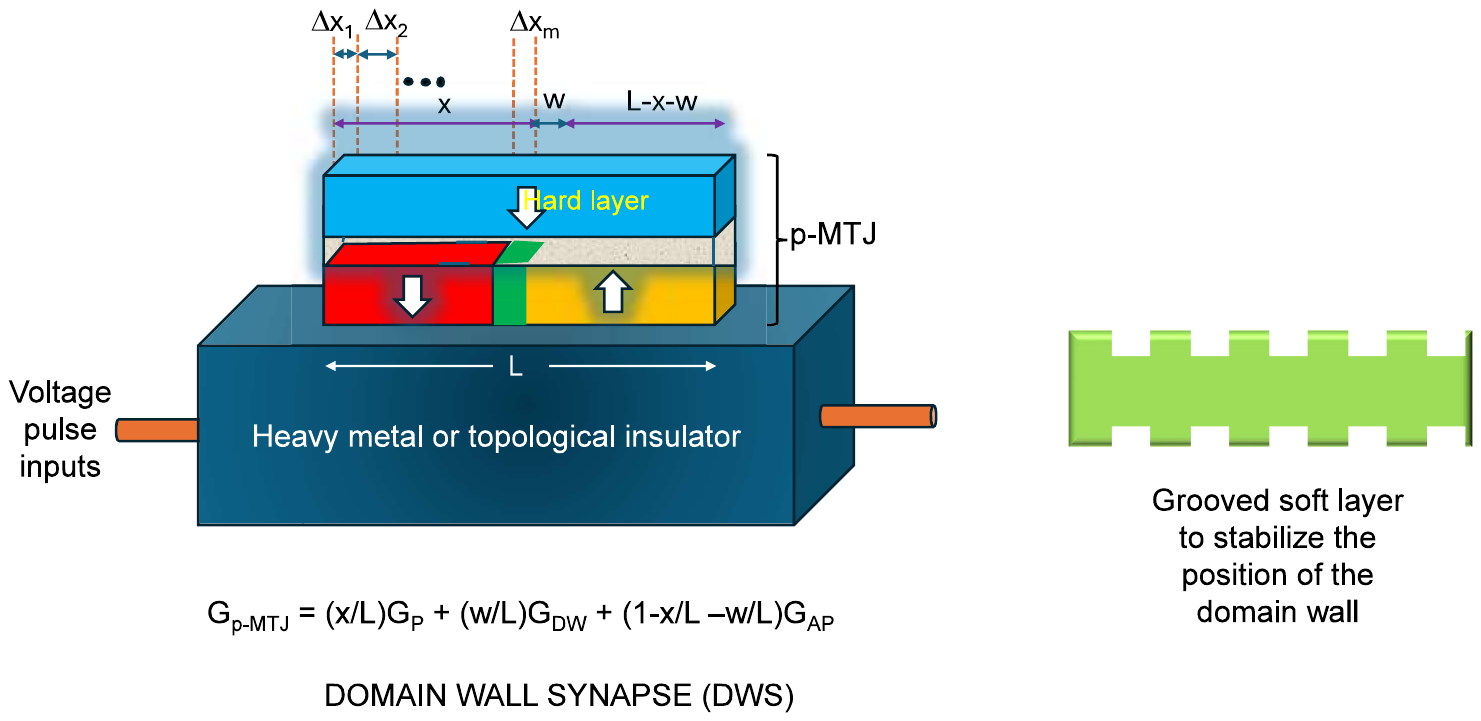}
\vspace{-1in}
\caption{\label{fig:DWS} A domain wall synapse. The domain wall position can be stabilized in the presence of thermal noise by using a periodically grooved soft layer}
\end{figure*}

We can update the weights by updating the resistance $r_i$. Weights can be implemented with memristors whose resistance can be changed/updated with a (gate) voltage, but a better option is a simple straintronic magnetic tunnel junction (s-MTJ) \cite{zhao} whose only difference with a generic magnetic tunnel junction is that the magnetization of the soft layer is rotated with electrically generated strain rather than the more common spin-transfer-torque, spin-orbit torque, or voltage-controlled-magnetic-anisotropy. The structure of a s-MTJ is shown in Fig. \ref{fig:s-MTJ}. The reason the s-MTJ is superior is that its transfer characteristic has a linear region where the conductance is {\it linearly proportional} to the gate voltage \cite{ieee,jphysd} (see Fig. \ref{fig:transfer} for the computed transfer characteristic of a s-MTJ). 
Ref. \cite{ieee} provided an analytical proof that such a linear region must exist and that its width is proportional to the antiparallel resistance of the s-MTJ. This linearity is very desirable and not easily available in any other device such as a memristor. The linearity allows more accurate control of the conductance where equal increments or decrements of the gate voltage will result in equal increase (long term potentiation) or decrease (long term depression) of the conductance improving accuracy in classification tasks \cite{debanjan1,debanjan2}. It may also benefit adaptive synpatogenesis.  

\subsection{Non-volatile weights}

Unlike memristor based weights, weights implemented with s-MTJs may appear to be {\it volatile} since the gate voltage should be kept on to maintain the conductance/resistance of the s-MTJ at a particular value. However, it must be noted that all that the gate voltage does is produce a strain in the s-MTJ's soft layer. If the produced strain survives after the gate voltage is removed, then the gate voltage need not be kept on and the weight would be non-volatile. This obviously requires {\it non-volatile strain}. Non-volatile strain induced by an electrical voltage in a piezoelectric has been reported by a number of authors \cite{chen,yang,wu1,wu2}. At this time, however, the robustness of non-volatile strain is questionable, i.e., whether it survives thermal recycling, and the retention time is undetermined, but the fact that remanent strain is found after withdrawal of the voltage that induces strain in a piezoelectric is a very encouraging development which portends {\it non-volatile weights} when they are implemented with s-MTJs.

\subsection{A device to measure the firing rate of a neuron}
 To measure the firing rate of the neuron, we employ a domain wall synapse (DWS) \cite{sengupta,incorvia} shown in Fig. \ref{fig:DWS}. It consists of a p-MTJ where the ferromagnetic hard and soft layers have perpendicular magnetic anisotropy. The p-MTJ is  fabricated on a heavy metal layer or a topological insulator. The output voltage pulses generated by a firing neuron are converted to current pulses and injected into the heavy metal layer or topological insulator to create successive pulses of spin-orbit torque that move the domain wall in the soft layer of the p-MTJ by discrete amounts $\Delta x_n$.

The conductance of the p-MTJ is given by \cite{sengupta}
\begin{equation}
    G_{p-MTJ} = \frac{x}{L}G_P
    + \frac{W}{L} G_{DW} + \left ( 1 - {{x+W}\over{L}} \right )G_{AP},
\end{equation}
where $G_P$ is the parallel conductance and $G_{AP}$ is the antiparallel conductance of the p-MTJ, $G_{DW}$ is the conductance associated with the domain wall, $L$ is the length of the soft layer,  and $x = \sum_i^m \Delta x_i$, the total domain wall displacement after $m$ pulses.\\

 \begin{figure}[!hbt]
 \vspace{-0.6in}
\centering
\includegraphics[width=0.4\textwidth,angle=270]{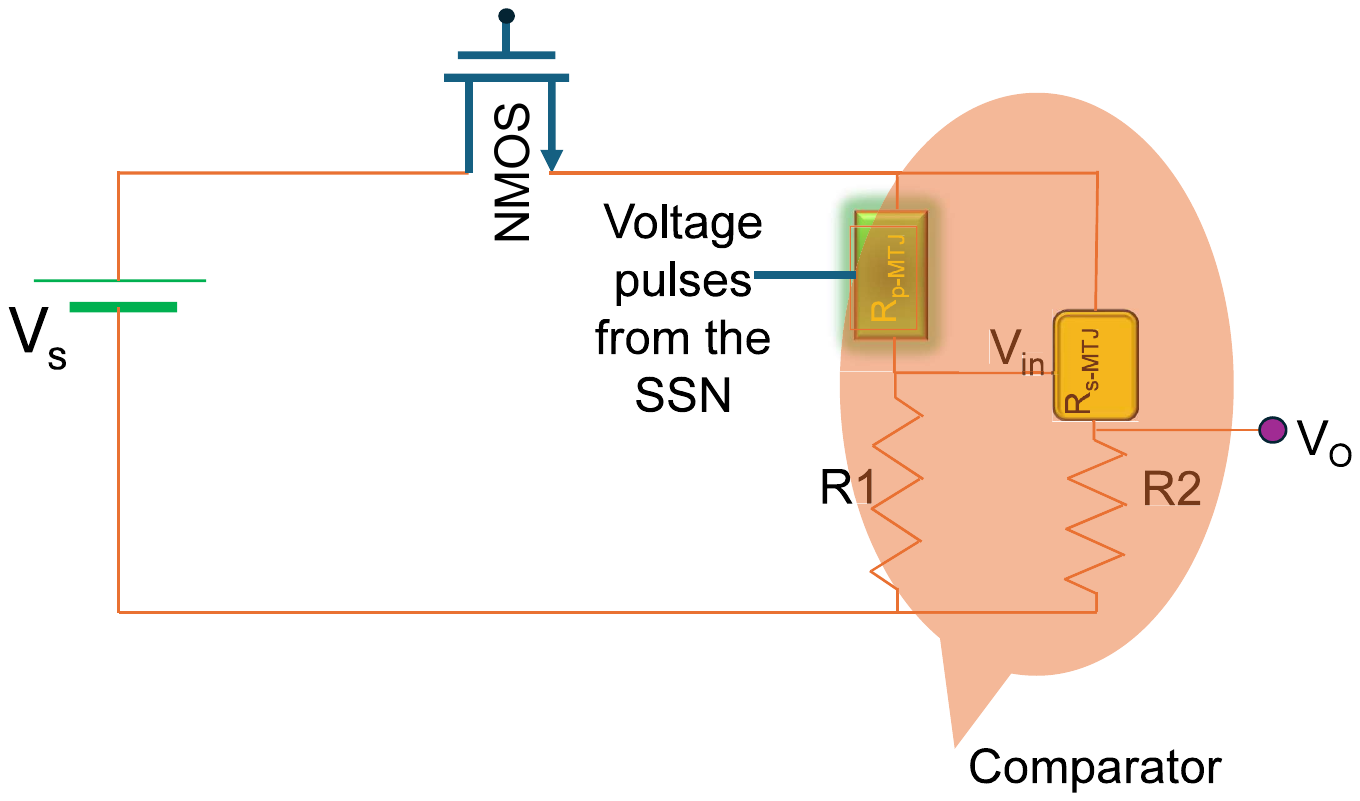}
\vspace{-0.6in}
\caption{\label{fig:circuit} Circuit to measure the firing rate and compare it with a threshold.}
\end{figure}

In order to measure the firing rate, we build the circuit of Fig. \ref{fig:circuit} where the DWS is placed in series with a resistor $R1$ and connected to a power supply voltage $V_s$ through an NMOS transistor. The gate voltage of the NMOS is turned on for a time duration $T$ and let us assume that the SSN fires $m$ times during that period. Because of slight dipole coupling between the hard and the soft layers of the p-MTJ, it will initially be in the antiparallel state. After time $T$, the conductance of the p-MTJ will be 
\begin{eqnarray}
    G_{p-MTJ} & = & {{\sum_i^m \Delta x_i}\over{L}}G_P
    + \frac{W}{L} G_{DW} \nonumber \\
    && + \left ( 1 - {{\sum_i^m \Delta x_i+W}\over{L}} \right ) G_{AP}\nonumber \\
   & = & m\frac{\Delta x}{L} G_P
   + \frac{W}{L} G_{DW} + \left ( 1 - {{\Delta x +W}\over{L}} \right ) G_{AP} \nonumber \\
   & = & m \frac{\Delta x}{L} \left ( G_P - G_{AP} \right )
    + \frac{W}{L} G_{DW} \nonumber \\
    && + \left (1 - \frac{W}{L} \right ) G_{AP},
    \label{rate}
\end{eqnarray}
where $\Delta x$ is the average displacement of the domain wall after a pulse. It can be determined experimentally with a magnetic force microscope after the soft layer has been subjected to a number of pulses.
Therefore, 
\begin{eqnarray}
    m & = & \frac{L}{\Delta x} \frac{1}{\left ( G_p - G_{AP} \right )} \times \nonumber \\
    && \left [ G_{p-MTJ} - \frac{W}{L} G_{DW} - G_{AP} \left (1 - \frac{W}{L} \right ) \right ].
\end{eqnarray}

\begin{figure*}[!hbt]
\centering
\includegraphics[width=0.9\textwidth]{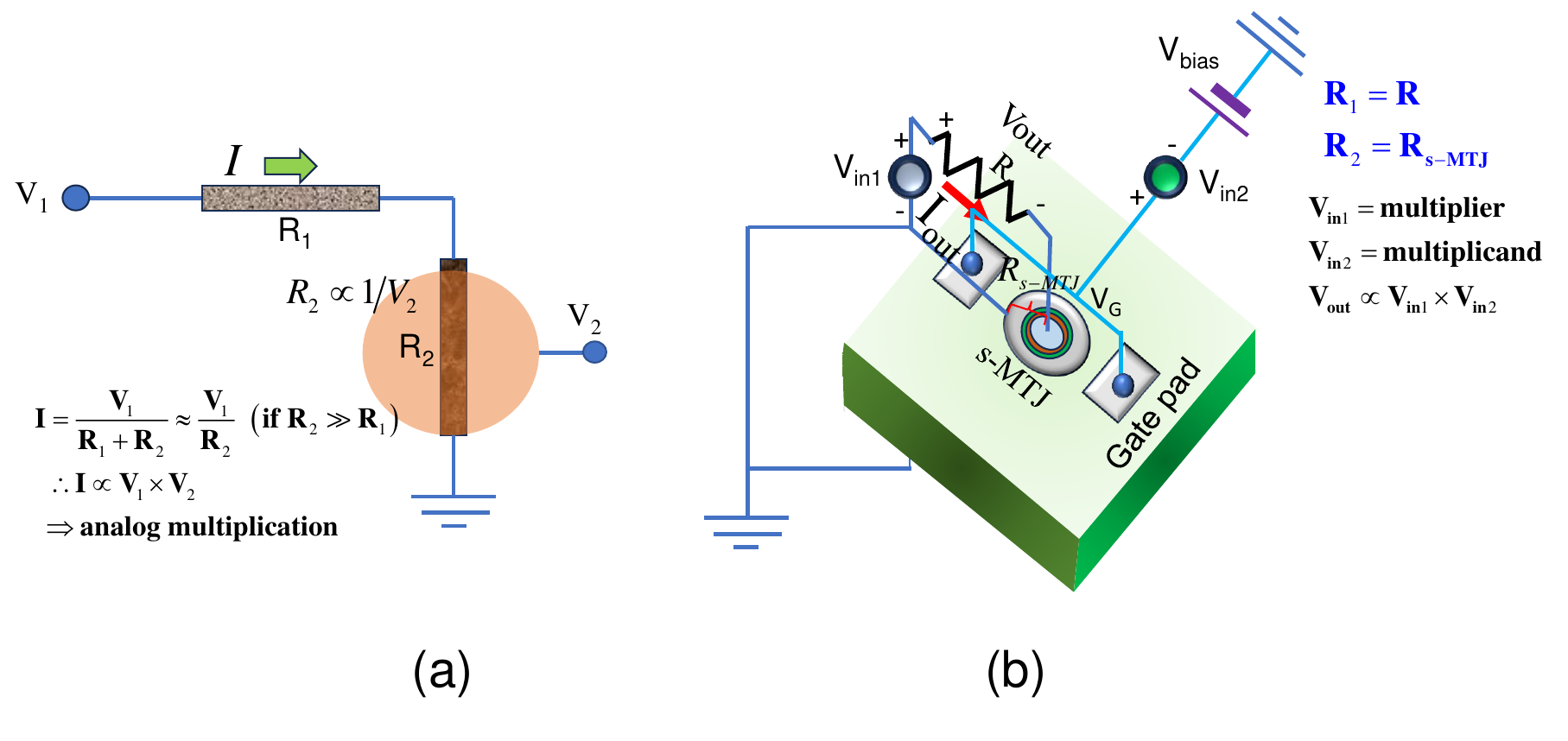}
\caption{\label{fig:analog-multiplier} (a) A voltage divider with a voltage tunable resistor can implement an {\bf analog multiplier}. (b) An actual implementation of an analog multiplier with a s-MTJ biased in the linear region of the transfer characteristic and some bias sources. Reproduced from \cite{mag} with permission of the Institute of Electrical and Electronics Engineers.}
\end{figure*}

\begin{figure}[!hbt]
\centering
\includegraphics[width=0.46\textwidth]{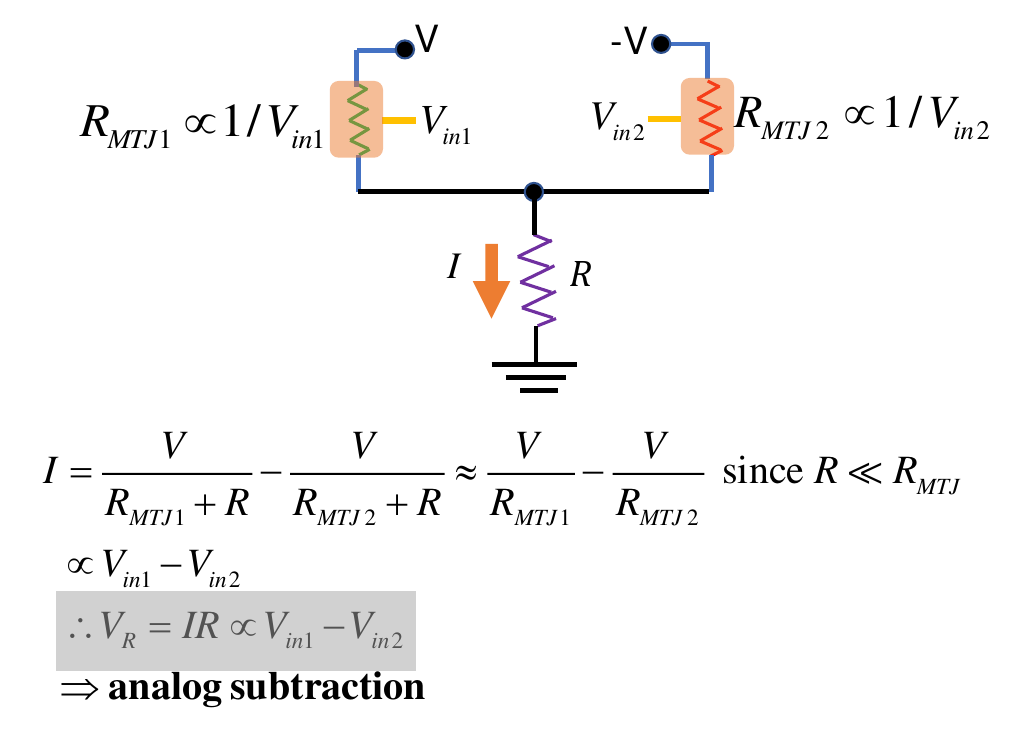}
\caption{\label{fig:subtractor} An {\bf analog subtractor} implemented with two s-MTJs acting as resistors whose resistances are inversely proportional to the gate voltages applied to them.}
\end{figure}

The quantities $W$, $L$, $G_P$, $G_{AP}$, $G_{DW}$  and $\Delta x$ are known. Hence by measuring $G_{p-MTJ}$, one can deduce the value of $m$. The firing rate is simply $m/T$. Thus, one can measure the firing rate.

\subsection{Determining if the firing rate exceeds a threshold - the comparator}

From Equation (\ref{rate}), we see that the p-MTJ conductance is linearly proportional to $m$ and hence to the firing rate $m/T$. Let the p-MTJ conductance corresponding to a threshold firing rate be $G_{p-MTJ}^T$ and its reciprocal be the resistance 
$R_{p-MTJ}^T$. In the voltage divider circuit of Fig. \ref{fig:circuit}, the voltage $V_{in}$ is $\frac{R1}{R1 + R_{p-MTJ}}$. At the threshold value,
\begin{equation}
    V_{in}^T = {{R1}\over{R1 +R_{p-MTJ}^T }}V_s.
\end{equation}

In Fig. \ref{fig:circuit}, the s-MTJ is a large cross-section straintronic MTJ whose resistance switches somewhat abruptly from high to low when the voltage applied to its gate exceeds a threshold value \cite{zhao}. The threshold depends on the energy barrier within the soft layer of the s-MTJ and hence can be engineered by engineering the shape of the s-MTJ (the ellipticity of the soft layer). We design the s-MTJ such that the threshold voltage for switching is equal to $V_{in}^T$. Then, if the firing rate of the SSN exceeds the threshold, the s-MTJ resistance will switch from high to low. Since $V_O = R2/\left (R2 + R_{s-MTJ} \right) $, a high value of $V_O$ will imply that the threshold firing rate has been exceeded, and a low value will indicate that the threshold has not been reached. Thus, by monitoring $V_O$, we can infer if the firing rate has exceeded a threshold.  This is the basis of the comparator.

\subsection{Analog multiplier}

An analog multiplier can be realized with a simple voltage divider network consisting of two resistors, one of whose resistance can be tuned with a gate voltage such that resistance is inversely proportional to the gate voltage. The implementation is shown in Fig. \ref{fig:analog-multiplier}. The voltage-dependent resistor $R_2$ is obviously implemented with a s-MTJ biased in the linear region of the transfer characteristic shown in Fig. \ref{fig:transfer} where the conductance is proportional to the applied gate voltage and hence the resistance is inversely proportional to the gate voltage. An exact implementation of the analog multiplier can be found in Ref. \cite{ieee,mag} and hence not repeated here.

\subsection{Analog subtractor}

An analog subtractor can also be realized with two voltage divider networks where once again we use s-MTJs to implement resistors whose resistances are inversely proportional to the gate voltage. We show only the circuit representations in Fig. \ref{fig:subtractor} since the device implementation follows trivially from there.

\subsection{A probability generator randomly generating binary bits 0 and 1 with tunable distribution of the probabilities of either bit}

This device can be realized with a simple binary stochastic neuron (BSN) implemented with a low barrier nanomagnet. The low barrier nanomagnet forms the soft layer of an MTJ whose resistance encodes the magnetization orientation of the soft layer. The resistance is given by
\begin{equation}
 R_{MTJ} = R_P + {{R_{AP} - R_P}\over{2}} [ 1 - cos \theta]   ,
\end{equation}
where $R_{P(AP)}$ is the parallel (antiparallel) resistance of the MTJ and $\theta$ is the angle between the magnetizations of the hard and the soft layer. If the magnetization of the soft layer subtends an acute angle with that of the hard layer, then the resistance is low and is interpreted as bit 0, whereas if the angle is obtuse, the resistance is high and is interpreted as bit 1. The probability of obtaining the bit zero can be tuned by injecting a spin polarized current into soft layer with varying spin polarization.

Let us say that the unit vector along the magnetization of the hard layer is $\hat{m}$ and that along the spin polarization is $\hat{s}$. If the scalar product $\hat{m} \cdot \hat{s}$ is positive, we will call the spin polarized current positive; otherwise, we will consider it as negative. The probability of the measured bit being zero depends on the sign and magnitude of the spin polarized current 
\section{Conclusion}
Adaptive synaptogenesis has been discussed as a neuromorphic learning approach that mitigates the challenge of catastrophic forgetfulness inherent to conventional ANN retraining. Rather than overwriting all past memory in a single disruptive event, adaptive synaptogenesis enables constructive learning over time and efficient long-term memory retention. Adaptively and continually adding and culling synaptic connections in an artificial neural network retains long-term learning gleaned through longitudinal experience. Conventional, global retraining approaches are inherently incapable of such comprehensive subjective experiential learning that encompasses even temporally distant or sparse events. The adaptive synaptogenesis approach is particularly relevant for edge sensing applications, in which data connectivity, bandwidth limitations, or computing resource constraints make complete periodic retraining impractical or impossible.

Simulation models demonstrated the functional viability of this construct, both in supervised and unsupervised modes of learning. Model inputs correlated with classes were automatically encoded through the creation of new synaptic connections and unused or spurious connections were shed for computational efficiency. Learning accuracy for adaptive synaptogenesis modes were comparable to kNN in a series of benchmark tests.

Beyond the theoretical construct and modeling of this architecture, a hardware implementation has been proposed using power-efficient nanomagnetic devices, capable of variable output that can be changed within a nanosecond. These straintronic spin neurons use linearly gate-regulated channel conductance to encode weights in memory. While nonvolatility of strain state has been reported, more work is required to understand the environmental limits. Furthermore, several logic devices have been described as functional elements of an ANN with adaptive synaptogenesis capability, including a firing rate comparator, multiplier, and subtractor.

In future work, the adaptive synaptogenesis model will be implemented using straintronic magnetic tunnel junction devices and characterized for comparison to the simulation results reported herein. Subsequently, the hardware will be used for real-time learning applications, such as embedded classification techniques in sensing applications and fluid neuromuscular control in robotic applications.
\begin{acknowledgments}
The authors are indebted to our late colleague Prof. William Levy of the University of Virginia for many illuminating discussions pertaining to neuroscience. This project was funded in part by the US National Science Foundation IUCRC Multifunctional Integrated System Technology Center at University of Virginia, US Air Force Office of Scientific Research under grant
FA865123CA023 given to the iDISPLA Convergence Lab at Virginia Commonwealth University and by the Virginia Innovation Partnership Corporation grant CCF23-0114-HE. 
\end{acknowledgments}
\bibliography{SP}

\begin{thebibliography}{22}%
\makeatletter
\providecommand \@ifxundefined [1]{%
 \@ifx{#1\undefined}
}%
\providecommand \@ifnum [1]{%
 \ifnum #1\expandafter \@firstoftwo
 \else \expandafter \@secondoftwo
 \fi
}%
\providecommand \@ifx [1]{%
 \ifx #1\expandafter \@firstoftwo
 \else \expandafter \@secondoftwo
 \fi
}%
\providecommand \natexlab [1]{#1}%
\providecommand \enquote  [1]{``#1''}%
\providecommand \bibnamefont  [1]{#1}%
\providecommand \bibfnamefont [1]{#1}%
\providecommand \citenamefont [1]{#1}%
\providecommand \href@noop [0]{\@secondoftwo}%
\providecommand \href [0]{\begingroup \@sanitize@url \@href}%
\providecommand \@href[1]{\@@startlink{#1}\@@href}%
\providecommand \@@href[1]{\endgroup#1\@@endlink}%
\providecommand \@sanitize@url [0]{\catcode `\\12\catcode `\$12\catcode `\&12\catcode `\#12\catcode `\^12\catcode `\_12\catcode `\%12\relax}%
\providecommand \@@startlink[1]{}%
\providecommand \@@endlink[0]{}%
\providecommand \url  [0]{\begingroup\@sanitize@url \@url }%
\providecommand \@url [1]{\endgroup\@href {#1}{\urlprefix }}%
\providecommand \urlprefix  [0]{URL }%
\providecommand \Eprint [0]{\href }%
\providecommand \doibase [0]{https://doi.org/}%
\providecommand \selectlanguage [0]{\@gobble}%
\providecommand \bibinfo  [0]{\@secondoftwo}%
\providecommand \bibfield  [0]{\@secondoftwo}%
\providecommand \translation [1]{[#1]}%
\providecommand \BibitemOpen [0]{}%
\providecommand \bibitemStop [0]{}%
\providecommand \bibitemNoStop [0]{.\EOS\space}%
\providecommand \EOS [0]{\spacefactor3000\relax}%
\providecommand \BibitemShut  [1]{\csname bibitem#1\endcsname}%
\let\auto@bib@innerbib\@empty
\bibitem [{\citenamefont {Feder}\ \emph {et~al.}(1999)\citenamefont {Feder}, \citenamefont {Leonard},\ and\ \citenamefont {Smith}}]{Hans}%
  \BibitemOpen
  \bibfield  {author} {\bibinfo {author} {\bibfnamefont {H.~J.~S.}\ \bibnamefont {Feder}}, \bibinfo {author} {\bibfnamefont {J.~J.}\ \bibnamefont {Leonard}},\ and\ \bibinfo {author} {\bibfnamefont {C.~M.}\ \bibnamefont {Smith}},\ }\bibfield  {title} {\bibinfo {title} {Adaptive mobile robot navigation and mapping},\ }\href {https://doi.org/10.1177/02783649922066484} {\bibfield  {journal} {\bibinfo  {journal} {The International Journal of Robotics Research}\ }\textbf {\bibinfo {volume} {18}},\ \bibinfo {pages} {650} (\bibinfo {year} {1999})}\BibitemShut {NoStop}%
\bibitem [{\citenamefont {Mei}\ \emph {et~al.}(2006)\citenamefont {Mei}, \citenamefont {Lu}, \citenamefont {Hu},\ and\ \citenamefont {Lee}}]{Mei}%
  \BibitemOpen
  \bibfield  {author} {\bibinfo {author} {\bibfnamefont {Y.}~\bibnamefont {Mei}}, \bibinfo {author} {\bibfnamefont {Y.-H.}\ \bibnamefont {Lu}}, \bibinfo {author} {\bibfnamefont {Y.}~\bibnamefont {Hu}},\ and\ \bibinfo {author} {\bibfnamefont {C.}~\bibnamefont {Lee}},\ }\bibfield  {title} {\bibinfo {title} {Deployment of mobile robots with energy and timing constraints},\ }\href {https://doi.org/10.1109/TRO.2006.875494} {\bibfield  {journal} {\bibinfo  {journal} {IEEE Transactions on Robotics}\ }\textbf {\bibinfo {volume} {22}},\ \bibinfo {pages} {507} (\bibinfo {year} {2006})}\BibitemShut {NoStop}%
\bibitem [{\citenamefont {Lluvia}\ \emph {et~al.}(2021)\citenamefont {Lluvia}, \citenamefont {Lazkano},\ and\ \citenamefont {Ansuategi}}]{Lluvia}%
  \BibitemOpen
  \bibfield  {author} {\bibinfo {author} {\bibfnamefont {I.}~\bibnamefont {Lluvia}}, \bibinfo {author} {\bibfnamefont {E.}~\bibnamefont {Lazkano}},\ and\ \bibinfo {author} {\bibfnamefont {A.}~\bibnamefont {Ansuategi}},\ }\bibfield  {title} {\bibinfo {title} {Active mapping and robot exploration: A survey},\ }\bibfield  {journal} {\bibinfo  {journal} {Sensors}\ }\textbf {\bibinfo {volume} {21}},\ \href {https://doi.org/10.3390/s21072445} {10.3390/s21072445} (\bibinfo {year} {2021})\BibitemShut {NoStop}%
\bibitem [{\citenamefont {Mermillod}\ \emph {et~al.}(2013)\citenamefont {Mermillod}, \citenamefont {Bugaiska},\ and\ \citenamefont {BONIN}}]{Mermillod}%
  \BibitemOpen
  \bibfield  {author} {\bibinfo {author} {\bibfnamefont {M.}~\bibnamefont {Mermillod}}, \bibinfo {author} {\bibfnamefont {A.}~\bibnamefont {Bugaiska}},\ and\ \bibinfo {author} {\bibfnamefont {P.}~\bibnamefont {BONIN}},\ }\bibfield  {title} {\bibinfo {title} {The stability-plasticity dilemma: investigating the continuum from catastrophic forgetting to age-limited learning effects},\ }\bibfield  {journal} {\bibinfo  {journal} {Frontiers in Psychology}\ }\textbf {\bibinfo {volume} {4}},\ \href {https://doi.org/10.3389/fpsyg.2013.00504} {10.3389/fpsyg.2013.00504} (\bibinfo {year} {2013})\BibitemShut {NoStop}%
\bibitem [{\citenamefont {Hajizada}\ \emph {et~al.}(2022)\citenamefont {Hajizada}, \citenamefont {Berggold}, \citenamefont {Iacono}, \citenamefont {Glover},\ and\ \citenamefont {Sandamirskaya}}]{hajizada}%
  \BibitemOpen
  \bibfield  {author} {\bibinfo {author} {\bibfnamefont {E.}~\bibnamefont {Hajizada}}, \bibinfo {author} {\bibfnamefont {P.}~\bibnamefont {Berggold}}, \bibinfo {author} {\bibfnamefont {M.}~\bibnamefont {Iacono}}, \bibinfo {author} {\bibfnamefont {A.}~\bibnamefont {Glover}},\ and\ \bibinfo {author} {\bibfnamefont {Y.}~\bibnamefont {Sandamirskaya}},\ }\bibfield  {title} {\bibinfo {title} {Interactive continual learning for robots: a neuromorphic approach},\ }in\ \href {https://doi.org/10.1145/3546790.3546791} {\emph {\bibinfo {booktitle} {International Conference on Neuromorphic Systems 2022}}},\ \bibinfo {series and number} {ICONS '22}\ (\bibinfo  {publisher} {ACM},\ \bibinfo {address} {New York, NY, USA},\ \bibinfo {year} {2022})\BibitemShut {NoStop}%
\bibitem [{\citenamefont {Levy}(1996)}]{Levy1996}%
  \BibitemOpen
  \bibfield  {author} {\bibinfo {author} {\bibfnamefont {W.~B.}\ \bibnamefont {Levy}},\ }\bibfield  {title} {{\selectlanguage {English}\bibinfo {title} {A sequence predicting ca3 is a flexible associator that learns and uses context to solve hippocampal-like tasks}},\ }\href {https://doi.org/10.1002/(SICI)1098-1063(1996)6:6<579::AID-HIPO3>3.0.CO;2-C} {\bibfield  {journal} {\bibinfo  {journal} {Hippocampus}\ }\textbf {\bibinfo {volume} {6}},\ \bibinfo {pages} {579} (\bibinfo {year} {1996})}\BibitemShut {NoStop}%
\bibitem [{\citenamefont {August}\ and\ \citenamefont {Levy}(1999)}]{August1999}%
  \BibitemOpen
  \bibfield  {author} {\bibinfo {author} {\bibfnamefont {D.~A.}\ \bibnamefont {August}}\ and\ \bibinfo {author} {\bibfnamefont {W.~B.}\ \bibnamefont {Levy}},\ }\bibfield  {title} {{\selectlanguage {English}\bibinfo {title} {Temporal sequence compression by an integrate-and-fire model of hippocampal area ca3}},\ }\href {https://doi.org/10.1023/a:1008861001091} {\bibfield  {journal} {\bibinfo  {journal} {Journal of Computational Neuroscience}\ }\textbf {\bibinfo {volume} {6}},\ \bibinfo {pages} {71} (\bibinfo {year} {1999})}\BibitemShut {NoStop}%
\bibitem [{\citenamefont {Hendrycks}\ and\ \citenamefont {Gimpel}(2016)}]{hendrycks2016gelu}%
  \BibitemOpen
  \bibfield  {author} {\bibinfo {author} {\bibfnamefont {D.}~\bibnamefont {Hendrycks}}\ and\ \bibinfo {author} {\bibfnamefont {K.}~\bibnamefont {Gimpel}},\ }\bibfield  {title} {\bibinfo {title} {Gaussian error linear units (gelus)},\ }\href@noop {} {\bibfield  {journal} {\bibinfo  {journal} {arXiv preprint arXiv:1606.08415}\ } (\bibinfo {year} {2016})}\BibitemShut {NoStop}%
\bibitem [{\citenamefont {Baxter}\ and\ \citenamefont {Levy}(2020)}]{Baxter2020ConstructingMechanisms}%
  \BibitemOpen
  \bibfield  {author} {\bibinfo {author} {\bibfnamefont {R.~A.}\ \bibnamefont {Baxter}}\ and\ \bibinfo {author} {\bibfnamefont {W.~B.}\ \bibnamefont {Levy}},\ }\bibfield  {title} {\bibinfo {title} {{Constructing multilayered neural networks with sparse, data-driven connectivity using biologically-inspired, complementary, homeostatic mechanisms}},\ }\href {https://doi.org/10.1016/j.neunet.2019.09.025} {\bibfield  {journal} {\bibinfo  {journal} {Neural Networks}\ }\textbf {\bibinfo {volume} {122}},\ \bibinfo {pages} {68} (\bibinfo {year} {2020})}\BibitemShut {NoStop}%
\bibitem [{\citenamefont {Biswas}\ \emph {et~al.}(2015)\citenamefont {Biswas}, \citenamefont {Atulasimha},\ and\ \citenamefont {Bandyopadhyay}}]{ayan}%
  \BibitemOpen
  \bibfield  {author} {\bibinfo {author} {\bibfnamefont {A.~K.}\ \bibnamefont {Biswas}}, \bibinfo {author} {\bibfnamefont {J.}~\bibnamefont {Atulasimha}},\ and\ \bibinfo {author} {\bibfnamefont {S.}~\bibnamefont {Bandyopadhyay}},\ }\bibfield  {title} {\bibinfo {title} {The straintronic spin neuron},\ }\href@noop {} {\bibfield  {journal} {\bibinfo  {journal} {Nanotechnology}\ }\textbf {\bibinfo {volume} {26}},\ \bibinfo {pages} {285201} (\bibinfo {year} {2015})}\BibitemShut {NoStop}%
\bibitem [{\citenamefont {Zhao}\ \emph {et~al.}(2016)\citenamefont {Zhao}, \citenamefont {Jamali}, \citenamefont {D’Souza}, \citenamefont {Zhang}, \citenamefont {Bandyopadhyay}, \citenamefont {Atulasimha},\ and\ \citenamefont {Wang}}]{zhao}%
  \BibitemOpen
  \bibfield  {author} {\bibinfo {author} {\bibfnamefont {Z.~Y.}\ \bibnamefont {Zhao}}, \bibinfo {author} {\bibfnamefont {M.}~\bibnamefont {Jamali}}, \bibinfo {author} {\bibfnamefont {N.}~\bibnamefont {D’Souza}}, \bibinfo {author} {\bibfnamefont {D.}~\bibnamefont {Zhang}}, \bibinfo {author} {\bibfnamefont {S.}~\bibnamefont {Bandyopadhyay}}, \bibinfo {author} {\bibfnamefont {J.}~\bibnamefont {Atulasimha}},\ and\ \bibinfo {author} {\bibfnamefont {J.~P.}\ \bibnamefont {Wang}},\ }\bibfield  {title} {\bibinfo {title} {Giant voltage manipulation of mgo-based magnetic tunnel junctions via localized anisotropic strain: A potential pathway to ultra-energy-efficient memory technology},\ }\href@noop {} {\bibfield  {journal} {\bibinfo  {journal} {Appl. Phys. Lett.}\ }\textbf {\bibinfo {volume} {109}},\ \bibinfo {pages} {092403} (\bibinfo {year} {2016})}\BibitemShut {NoStop}%
\bibitem [{\citenamefont {Rahman}\ and\ \citenamefont {Bandyopadhyay}(2022)}]{ieee}%
  \BibitemOpen
  \bibfield  {author} {\bibinfo {author} {\bibfnamefont {R.}~\bibnamefont {Rahman}}\ and\ \bibinfo {author} {\bibfnamefont {S.}~\bibnamefont {Bandyopadhyay}},\ }\bibfield  {title} {\bibinfo {title} {A non-volatile all-spin non-binary matrix multiplier: An efficient hardware accelerator for machine learning},\ }\href@noop {} {\bibfield  {journal} {\bibinfo  {journal} {IEEE Trans. Elec. Dev.}\ }\textbf {\bibinfo {volume} {69}},\ \bibinfo {pages} {7120} (\bibinfo {year} {2022})}\BibitemShut {NoStop}%
\bibitem [{\citenamefont {Bandyopadhyay}(2025)}]{jphysd}%
  \BibitemOpen
  \bibfield  {author} {\bibinfo {author} {\bibfnamefont {S.}~\bibnamefont {Bandyopadhyay}},\ }\bibfield  {title} {\bibinfo {title} {Straintronic magnetic tunnel junctions for analog computation: A perspective},\ }\href@noop {} {\bibfield  {journal} {\bibinfo  {journal} {J. Phys. D: Appl. Phys}\ }\textbf {\bibinfo {volume} {58}},\ \bibinfo {pages} {152001} (\bibinfo {year} {2025})}\BibitemShut {NoStop}%
\bibitem [{\citenamefont {Yadav}\ \emph {et~al.}(2023)\citenamefont {Yadav}, \citenamefont {Gupta}, \citenamefont {Holla}, \citenamefont {Khan}, \citenamefont {Muduli},\ and\ \citenamefont {Bhowmik}}]{debanjan1}%
  \BibitemOpen
  \bibfield  {author} {\bibinfo {author} {\bibfnamefont {R.~S.}\ \bibnamefont {Yadav}}, \bibinfo {author} {\bibfnamefont {P.}~\bibnamefont {Gupta}}, \bibinfo {author} {\bibfnamefont {A.}~\bibnamefont {Holla}}, \bibinfo {author} {\bibfnamefont {K.~I.~A.}\ \bibnamefont {Khan}}, \bibinfo {author} {\bibfnamefont {P.~K.}\ \bibnamefont {Muduli}},\ and\ \bibinfo {author} {\bibfnamefont {D.}~\bibnamefont {Bhowmik}},\ }\bibfield  {title} {\bibinfo {title} {Demonstration of synaptic behavior in a heavy metal ferromagnetic metal oxide heterostructure based spintronic device for on-chip learning in crossbar-array-based neural networks},\ }\href@noop {} {\bibfield  {journal} {\bibinfo  {journal} {ACS Appl. Electron. Mater.}\ }\textbf {\bibinfo {volume} {5}},\ \bibinfo {pages} {484–97} (\bibinfo {year} {2023})}\BibitemShut {NoStop}%
\bibitem [{\citenamefont {Desai}\ \emph {et~al.}(2022)\citenamefont {Desai}, \citenamefont {Kaushik}, \citenamefont {Sharda},\ and\ \citenamefont {Bhowmik}}]{debanjan2}%
  \BibitemOpen
  \bibfield  {author} {\bibinfo {author} {\bibfnamefont {V.~B.}\ \bibnamefont {Desai}}, \bibinfo {author} {\bibfnamefont {D.}~\bibnamefont {Kaushik}}, \bibinfo {author} {\bibfnamefont {J.}~\bibnamefont {Sharda}},\ and\ \bibinfo {author} {\bibfnamefont {D.}~\bibnamefont {Bhowmik}},\ }\bibfield  {title} {\bibinfo {title} {On-chip learning of a domain-wall-synapse-crossbar-array-based convolutional neural network},\ }\href@noop {} {\bibfield  {journal} {\bibinfo  {journal} {Neuromorph. Comput. Eng.}\ }\textbf {\bibinfo {volume} {2}},\ \bibinfo {pages} {024006} (\bibinfo {year} {2022})}\BibitemShut {NoStop}%
\bibitem [{\citenamefont {Chen}\ and\ \citenamefont {et~al.}(2019)}]{chen}%
  \BibitemOpen
  \bibfield  {author} {\bibinfo {author} {\bibfnamefont {A.}~\bibnamefont {Chen}}\ and\ \bibinfo {author} {\bibnamefont {et~al.}},\ }\bibfield  {title} {\bibinfo {title} {Giant nonvolatile manipulation of magnetoresistance in magnetic tunnel junctions by electric fields via magnetoelectric coupling},\ }\href@noop {} {\bibfield  {journal} {\bibinfo  {journal} {Nat. Commun.}\ }\textbf {\bibinfo {volume} {10}},\ \bibinfo {pages} {243} (\bibinfo {year} {2019})}\BibitemShut {NoStop}%
\bibitem [{\citenamefont {Yang}\ \emph {et~al.}(2014)\citenamefont {Yang}, \citenamefont {Zhao}, \citenamefont {Zhang}, \citenamefont {Li}, \citenamefont {Gao}, \citenamefont {Yang}, \citenamefont {Hu}, \citenamefont {Miao}, \citenamefont {Liu}, \citenamefont {Chen}, \citenamefont {Nan},\ and\ \citenamefont {Gao}}]{yang}%
  \BibitemOpen
  \bibfield  {author} {\bibinfo {author} {\bibfnamefont {L.}~\bibnamefont {Yang}}, \bibinfo {author} {\bibfnamefont {Y.}~\bibnamefont {Zhao}}, \bibinfo {author} {\bibfnamefont {S.}~\bibnamefont {Zhang}}, \bibinfo {author} {\bibfnamefont {P.}~\bibnamefont {Li}}, \bibinfo {author} {\bibfnamefont {Y.}~\bibnamefont {Gao}}, \bibinfo {author} {\bibfnamefont {Y.}~\bibnamefont {Yang}}, \bibinfo {author} {\bibfnamefont {H.}~\bibnamefont {Hu}}, \bibinfo {author} {\bibfnamefont {P.}~\bibnamefont {Miao}}, \bibinfo {author} {\bibfnamefont {Y.}~\bibnamefont {Liu}}, \bibinfo {author} {\bibfnamefont {A.}~\bibnamefont {Chen}}, \bibinfo {author} {\bibfnamefont {C.~W.}\ \bibnamefont {Nan}},\ and\ \bibinfo {author} {\bibfnamefont {C.}~\bibnamefont {Gao}},\ }\bibfield  {title} {\bibinfo {title} {Bipolar loop-like non-volatile strain in the (001)-oriented pb(mg\(_{1/3}\)nb\(_{2/3}\))o\(_3\)-pbtio\(_3\) single crystals},\ }\href@noop {} {\bibfield  {journal} {\bibinfo  {journal} {Sci. Rep.}\ }\textbf {\bibinfo {volume} {4}},\
  \bibinfo {pages} {4591} (\bibinfo {year} {2014})}\BibitemShut {NoStop}%
\bibitem [{\citenamefont {Wu}\ and\ \citenamefont {et~al.}(2011{\natexlab{a}})}]{wu1}%
  \BibitemOpen
  \bibfield  {author} {\bibinfo {author} {\bibfnamefont {T.}~\bibnamefont {Wu}}\ and\ \bibinfo {author} {\bibnamefont {et~al.}},\ }\bibfield  {title} {\bibinfo {title} {Domain engineered switchable strain states in ferroelectric (011) [pb(mg\(_{1/3}\)nb\(_{2/3}\))o\(_3\)]\(_{1-x}\)-[pbtio3]\(_x\) (pmn-pt, x=0.32) single crystals},\ }\href@noop {} {\bibfield  {journal} {\bibinfo  {journal} {J. Appl Phys.}\ }\textbf {\bibinfo {volume} {109}},\ \bibinfo {pages} {124101} (\bibinfo {year} {2011}{\natexlab{a}})}\BibitemShut {NoStop}%
\bibitem [{\citenamefont {Wu}\ and\ \citenamefont {et~al.}(2011{\natexlab{b}})}]{wu2}%
  \BibitemOpen
  \bibfield  {author} {\bibinfo {author} {\bibfnamefont {T.}~\bibnamefont {Wu}}\ and\ \bibinfo {author} {\bibnamefont {et~al.}},\ }\bibfield  {title} {\bibinfo {title} {Electrical control of reversible and permanent magnetization reorientation for magnetoelectric memory devices},\ }\href@noop {} {\bibfield  {journal} {\bibinfo  {journal} {Appl. Phys. Lett.}\ }\textbf {\bibinfo {volume} {98}},\ \bibinfo {pages} {262504} (\bibinfo {year} {2011}{\natexlab{b}})}\BibitemShut {NoStop}%
\bibitem [{\citenamefont {Sengupta}\ \emph {et~al.}(2016)\citenamefont {Sengupta}, \citenamefont {Shim},\ and\ \citenamefont {Roy}}]{sengupta}%
  \BibitemOpen
  \bibfield  {author} {\bibinfo {author} {\bibfnamefont {A.}~\bibnamefont {Sengupta}}, \bibinfo {author} {\bibfnamefont {Y.}~\bibnamefont {Shim}},\ and\ \bibinfo {author} {\bibfnamefont {K.}~\bibnamefont {Roy}},\ }\bibfield  {title} {\bibinfo {title} {Proposal for all-spin artificial neural network: Emulating neural and synaptic functionalities through domain wall motion in ferromagnets},\ }\href@noop {} {\bibfield  {journal} {\bibinfo  {journal} {IEEE Trans. Biomed. Circ. Syst.}\ }\textbf {\bibinfo {volume} {10}},\ \bibinfo {pages} {1152–1160} (\bibinfo {year} {2016})}\BibitemShut {NoStop}%
\bibitem [{\citenamefont {Liu}\ \emph {et~al.}(2021)\citenamefont {Liu}, \citenamefont {Xiao}, \citenamefont {Cui}, \citenamefont {Incorvia}, \citenamefont {Bennett},\ and\ \citenamefont {Marinella}}]{incorvia}%
  \BibitemOpen
  \bibfield  {author} {\bibinfo {author} {\bibfnamefont {S.}~\bibnamefont {Liu}}, \bibinfo {author} {\bibfnamefont {I.~P.}\ \bibnamefont {Xiao}}, \bibinfo {author} {\bibfnamefont {C.}~\bibnamefont {Cui}}, \bibinfo {author} {\bibfnamefont {J.~A.~C.}\ \bibnamefont {Incorvia}}, \bibinfo {author} {\bibfnamefont {C.~H.}\ \bibnamefont {Bennett}},\ and\ \bibinfo {author} {\bibfnamefont {M.~J.}\ \bibnamefont {Marinella}},\ }\bibfield  {title} {\bibinfo {title} {A domain wall magnetic tunnel junction artificial synapse with notched geometry for accurate and efficient training of deep neural networks},\ }\href@noop {} {\bibfield  {journal} {\bibinfo  {journal} {Appl. Phys. Lett.}\ }\textbf {\bibinfo {volume} {118}},\ \bibinfo {pages} {202405} (\bibinfo {year} {2021})}\BibitemShut {NoStop}%
\bibitem [{\citenamefont {Bandyopadhyay}(2024)}]{mag}%
  \BibitemOpen
  \bibfield  {author} {\bibinfo {author} {\bibfnamefont {S.}~\bibnamefont {Bandyopadhyay}},\ }\bibfield  {title} {\bibinfo {title} {Magnetic straintronics for ultra-energy-efficient unconventional computing: A review},\ }\href@noop {} {\bibfield  {journal} {\bibinfo  {journal} {IEEE Trans. Magn.}\ }\textbf {\bibinfo {volume} {60}},\ \bibinfo {pages} {4100110} (\bibinfo {year} {2024})}\BibitemShut {NoStop}%
\end{thebibliography}%

\end{document}



\title{(Supplemental Material)\\
Tripartite Phonon-Magnon-Plasmon Coupling, Parametric Amplification, and Formation of a Phonon-Magnon-Plasmon Polariton in a Two-Dimensional Periodic Array of Magnetostrictive/Plasmonic Bilayered Nanodots}
}

\author{Sreya Pal}
\affiliation{Department of Condensed Matter and Materials Physics, S. N. Bose National Centre for Basic Sciences, Block–JD, Sector–III, Salt Lake, Kolkata, 700106, India}

\author{Pratap Kumar Pal}
\affiliation{Department of Condensed Matter and Materials Physics, S. N. Bose National Centre for Basic Sciences, Block–JD, Sector–III, Salt Lake, Kolkata, 700106, India}

\author{Raisa Fabiha}
\affiliation{Department of Electrical and Computer Engineering,
Virginia Commonwealth University, Richmond, VA 23284, USA}

\author{Supriyo Bandyopadhyay}%
\email{sbandy@vcu.edu}
\affiliation{Department of Electrical and Computer Engineering,
Virginia Commonwealth University, Richmond, VA 23284, USA}

\author{Anjan Barman}%
 \email{abarman@bose.res.in}
\affiliation{Department of Condensed Matter and Materials Physics, S. N. Bose National Centre for Basic Sciences, Block–JD, Sector–III, Salt Lake, Kolkata, 700106, India}

\maketitle

\textcolor{black}{\section{Position dependence of the reflectivity spectra for plasmonic and non-plasmonic samples}
To ascertain that the observed differences between the
reflectivity spectra of the Al-containing and Al-lacking
samples are indeed intrinsic (i.e., due to the presence and
absence of plasmons), and not a mere artifact of extrinsic
sample differences such as unintentional variations in
size, shape, material composition, etc., we have measured
the reflectivity oscillations and the corresponding spectra
in four different regions in each wafer (one with Al
and the other without) by focusing the TR-MOKE laser
spot in those four different regions as shown in Fig.S1.}
\begin{figure*}[htbp!]
\centering
\includegraphics[width=13cm]{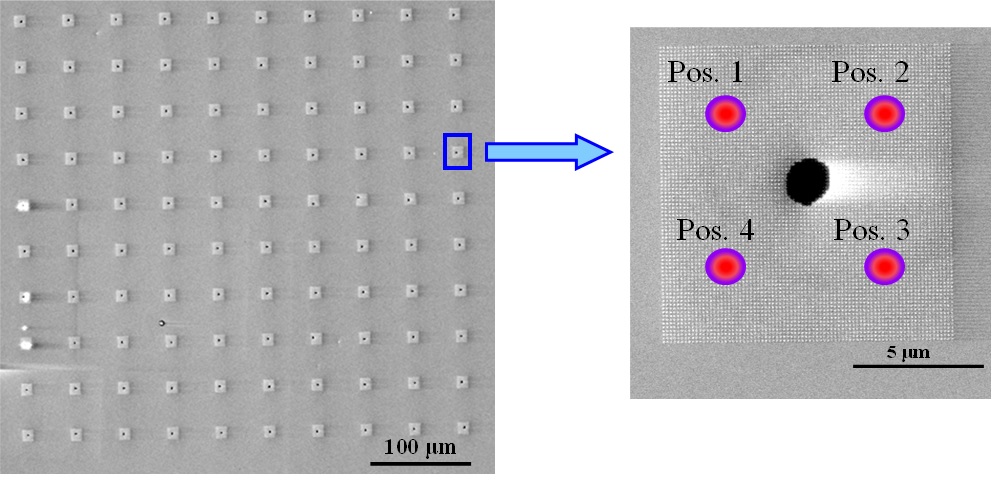}
\textcolor{black}{\caption{(a) Low resolution scanning electron microscope (SEM) images of a wafer showing an array of squares each containing
nanomagnets. (b) A higher resolution image of a randomly picked square showing the laser focused in four different regions. The laser spot covers approximately 36 nanomagnets. The black regions are alignment marks for navigation of the nanodot array across the wafer.}}
\label{FigS1_re}
\end{figure*}

\textcolor{black}{The results are presented in Fig. S2 and Fig. S3 for the Al-lacking and Al-containing samples respectively. The Al-lacking samples show the same behavior in all four probed regions (always 2 peaks in the spectra occurring at about the same frequencies) as do the Al-containing samples (always 4 peaks, again occurring at about the same frequencies). Their difference is remarkably consistent. This confirms that the observed differences in the spectra between the plasmonic and non-plasmonic specimens are intrinsic and not extrinsic.}

\begin{figure*}[htbp!]
\centering
\includegraphics[width=13cm]{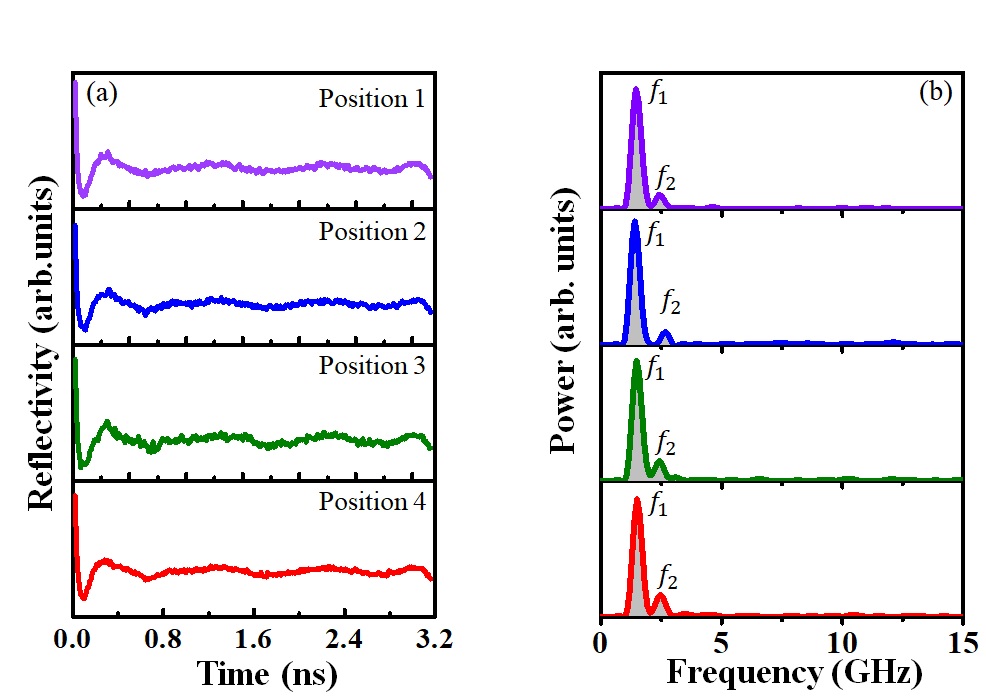}
\textcolor{black}{\caption{(a) Background-subtracted time-resolved data for the reflectivity and (b) the corresponding FFT in four “samples” devoid of Al. The four different “positions” refer to four different regions of the wafer, or effectively four different “samples” each consisting of $\sim$36 nanomagnets. The spectra are remarkably similar in all samples. There are always two peaks $(f\textsubscript{1})$ and $(f\textsubscript{2})$ that occur at about the same frequencies.}}
\label{FigS3}
\end{figure*}

\begin{figure*}[htbp!]
\centering
\includegraphics[width=12cm]{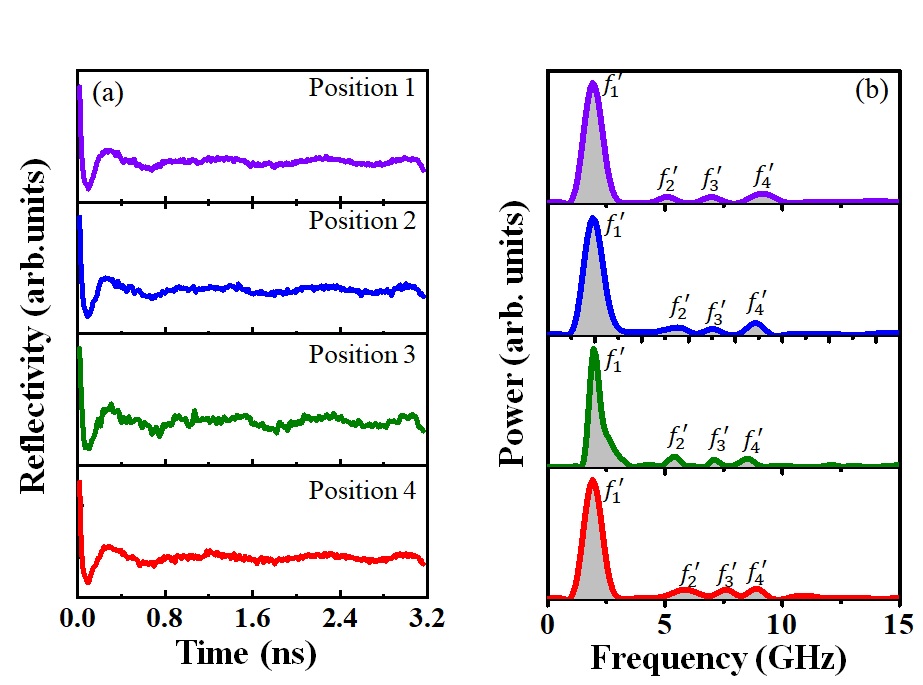}
\textcolor{black}{\caption{(a) Reflectivity oscillations and (b) the corresponding FFT in four 'samples' containing Al. The four different 'positions' refer to four
different regions of the wafer, or effectively four different 'samples' each consisting of $\sim$36 nanomagnets. The spectra are remarkably similar in all samples. There are always four peaks $f\textsubscript{1}^{'}$, $f\textsubscript{2}^{'}$, $f\textsubscript{3}^{'}$ and $f\textsubscript{4}^{'}$that occur at about the same frequencies.}}
\label{FigS4}
\end{figure*}

\newpage
\section {Spin-wave mode analysis for Co nanomagnets (Non-plasmonic samples}
To understand the characteristics of three magnetic modes in the samples without Al, including the Kittel-type and other modes, we used an in-house simulator Dotmag \cite{Kumar12} to calculate their spatial properties (power and phase). The results, presented in Fig. \ref{FigS1}, reveal distinct features.
\begin{figure*}[htbp!]
\centering
\includegraphics[width=13cm]{Figures/figure S1.jpg}
\caption{Simulated power and phase profiles of different spin-wave modes (mode 1, mode 2, mode 3) for Co nanomagnets without Al at four bias magnetic fields. Colour maps for power and phase distributions are shown below the images.}
\label{FigS1}
\end{figure*}
For mode '2', the power profile indicates a uniform distribution of spin-wave (SW) power at the nanomagnet's center, resembling a uniform precessional mode or center mode (CM). Despite high power at the edges, the central distribution is significant. In contrast, mode '1' shows power concentrated at the edges, termed the edge mode (EM), with power focused at two vertical edges perpendicular to the bias field. Mode '3' is akin to '2', except the quantization axis rotates about 45° with respect to the major axis of the elliptical nanomagnet. Analyzing phase profiles, we observe spins precessing alternately in opposite phases, creating azimuthal contrast with an azimuthal mode quantization number of 2. Notably, power distributions for these modes (both at the center and vertical edges of the nanomagnet) remain remarkably stable, showing minimal change with variations in external bias field strength.


\bibliography{SP}